# Vaccination and SARS-CoV-2 variants:

# how much containment is still needed? A quantitative assessment.

Giulia Giordano[1,*], Marta Colaneri[2], Alessandro Di Filippo[2], Franco Blanchini[3], Paolo Bolzern[4], Giuseppe De Nicolao[5], Paolo Sacchi[2], Raffaele Bruno[2,6,**], Patrizio Colaneri[4,7,**]

[1] Department of Industrial Engineering, University of Trento, Trento, Italy.

[2] Division of Infectious Diseases I, Fondazione IRCCS Policlinico San Matteo, Pavia, Italy.

[3] Dipartimento di Scienze Matematiche, Informatiche e Fisiche, University of Udine, Udine, Italy.

[4] Dipartimento di Elettronica, Informazione e Bioingegneria, Politecnico di Milano, Italy.

[5] Department of Electrical, Computer and Biomedical Engineering, University of Pavia, Pavia, Italy.

[6] Department of Clinical, Surgical, Diagnostic, and Paediatric Sciences, University of Pavia, Pavia, Italy.

[7] IEIIT-CNR, Milano, Italy.

* Corresponding author. E-mail: *giulia.giordano@unitn.it*.

** Co-senior authors.

February 15, 2021

**Despite the progress in medical care[1], combined population-wide interventions (such as physical distancing, testing and contact tracing) are still crucial to manage the SARS-CoV-2 pandemic, aggravated by the emergence of new highly transmissible variants. We combine the compartmental SIDARTHE model[2], predicting the course of COVID-19 infections, with a new data-based model that projects new cases onto casualties and healthcare system costs. Based on the Italian case study, we outline several scenarios: mass vaccination campaigns with different paces, different transmission rates due to new variants, and different enforced countermeasures, including the alternation of opening and closure phases. Our results demonstrate that non-pharmaceutical interventions (NPIs) have a higher impact on the epidemic evolution than vaccination, which advocates for the need to keep containment measures in place throughout the vaccination campaign. We also show that, if intermittent open-close strategies are adopted, deaths and healthcare system costs can be drastically reduced, without any aggravation of socioeconomic losses, as long as one has the foresight to start with a closing phase rather than an opening one.**

The infeasibility of long-term lockdowns (due to detrimental socioeconomic consequences and behavioural fatigue[3,4]) and of effective contact tracing at high case numbers led many countries to invest in mass vaccination to control the COVID-19 pandemic. Since the SARS-CoV-2 genome was sequenced in January 2020[5], researchers rushed to develop a vaccine suitable for mass distribution[6,7]. As of February 1st, 2021, 2 vaccines (Moderna and Pfizer/BioNTech) were approved by FDA for full use[8,9] and 8 more for limited use[10]. The registration clinical trials report 94% and 95% efficacy rates respectively for Moderna and Pfizer/BioNTech vaccines[11,12], with a favourable safety profile. Italy began its vaccination campaign on December 27th, 2020 with the delivery of the first 9,750 doses of Pfizer's BNT162b[13] destined for hospital healthcare workers. A vaccination program prioritizing all healthcare workers, nursing home residents and people over 80 years of age[14] was conducted starting from January 2021 with the delivery of further batches of the two approved vaccines[15,16]. Multi-pronged countermeasures, including distancing, testing and tracing, are necessary to achieve a sustained reduction in infection cases[17], even more so in the light of the recent emergence of new SARS-CoV-2 variants[18], such as B.1.1.7 and B1.351. These variants are reported to have an increased transmissibility[19,20] and possibly cause a more severe disease[21] compared to the original strain, although the efficacy of both currently authorised vaccines appears to be unaffected[22]. However, vaccination alone is not expected to control the spread of the infection: even under the optimistic assumption that no vaccine-resistant variants emerge, a carefully planned vaccination campaign[23,24] needs to be coordinated with NPIs[25] until sufficient coverage is reached.

In this complex layout, where vaccines and new variants are potential game-changers, models to forecast epidemic scenarios and assess the associated healthcare costs are essential. Our proposed integrated model (**Figure 1A**) feeds the predicted evolution of new positive cases, provided by the compartmental model SIDARTHE[2], extended to include the effect of vaccination, into a new data-based dynamic model, derived from Italian field data, that, once fed by the new cases, computes the time profile of the resulting healthcare system costs (hospital and ICU occupancy and deaths). To capture the progressive vaccination of the older population, the model takes into account the specific aggravation and death probability for different age classes (**Extended Data Figure 5**). Details are provided in the Methods.

We compare different scenarios to assess the effect of mass vaccination campaigns with different paces, in the presence of varying profiles of the reproduction number $\mathcal{R}_0$ over time, due to specific SARS-CoV-2 variants and/or restrictions. We consider four *effective vaccination* schedules (**Extended Data Figure 4**), obtained by modulating linearly the speed of the four phases T1-T4 of Italy's vaccination plan[26] so as to yield a different fraction of successfully immunised people within one year: absent (0%); slow (46%); medium (64%); fast (90%). We also consider five different profiles of $\mathcal{R}_0$ as a function of time: constant $\mathcal{R}_0 = 1.27$ (high transmission); Open-Close periodic $\mathcal{R}_0$ with average value 1.1, alternating first Openings ($\mathcal{R}_0 = 1.3$), and then Closures ($\mathcal{R}_0 = 0.9$); constant $\mathcal{R}_0 = 1.1$; Close-Open periodic $\mathcal{R}_0$ with average value 1.1, alternating first Closures and then Openings; constant $\mathcal{R}_0 = 0.9$ (eradication).

Our main findings and implications for policy-makers are outlined in Table 1 and summarised by the deaths vs. speed curves in **Figure 1B**, showing the death toll as a function of the vaccination speed for each $\mathcal{R}_0$ profile. The combination of the four vaccination schedules with the five $\mathcal{R}_0$ profiles leads to twenty distinct scenarios, associated with the dots in the deaths vs. speed curves (**Figure 1B**). Leaving aside eradication, associated with an almost constant curve (green), with all the other $\mathcal{R}_0$ profiles, vaccination saves at least 44% of lives compared to no vaccination, and even 50% and 57% with medium and fast schedules. The deaths vs. speed curves are flatter when $\mathcal{R}_0$ is kept smaller: stringent NPIs drastically reduce sensitivity to vaccination delays. Containment strategies have an impact on human losses that is 5-fold: depending on the $\mathcal{R}_0$ profile (not considering eradication), deaths in the period from February 2021 to January 2022 vary in the range of 39,000-196,000 (slow vaccination), 35,000-174,000 (medium vaccination) and 31,000-146,000 (fast vaccination). Therefore, NPIs have a much larger effect than vaccination speed. In the planning of mid-term interventions, pre-emption reduces life and healthcare system costs at no socio-economic cost: intermittent containment strategies with the same average $\mathcal{R}_0$ involve the same amount of socioeconomic restrictions, but starting with a Closing phase (Close-Open, purple curve) improves on constant containment (yellow), which is in turn better than starting with an Opening phase (Open-Close, orange). For all vaccination schedules, the Close-Open strategy saves no less than 23,000 lives compared to the Open-Close strategy. Hospital and ICU occupancy as a function of the vaccination speed follow a similar pattern (**Extended Data Figure 8**).

Considering a medium vaccination speed, **Figure 2** shows the epidemic evolution for different constant values of $\mathcal{R}_0$ (the scenarios in the absence of vaccination are in **Extended Data Figure 2**). In spite of vaccination and of containment measures, a higher transmissibility due to the spread of new variants would cause a dramatic surge in infection cases, leading to a peak of some 9,000 ICU beds needed and more than 1,600 daily deaths before sufficient population immunity can be reached. A significant reduction in hospital occupancy and death toll can be obtained by enforcing, throughout the vaccination campaign, NPIs to reduce $\mathcal{R}_0$, which need to be even more stringent in the presence of highly transmissible variants.

The need to enforce new restrictions is likely to trigger intermittent containment measures, with the alternation of higher-$\mathcal{R}_0$ and lower-$\mathcal{R}_0$ phases[27,28]. In Open-Close strategies, closures are delayed and only enforced when the pressure on the healthcare system becomes unbearable. Each intermittent Open-Close strategy can be associated with a Close-Open strategy that alternates the same opening and closing phases, with the only difference of starting with a closure. **Figure 3** compares the two different intermittent strategies, with average $\mathcal{R}_0$ equal to 1.1, under medium-speed vaccination (the comparison in the absence of vaccination is in **Extended Data Figure 3**). It appears that opening first (Open-Close) or closing first (Close-Open) strongly affects healthcare system costs (which depend on case numbers), while socioeconomic costs (which depend on the duration and stringency of restrictions) are substantially unchanged. Without aggravation of social and economic losses, a pre-emptive Close-Open strategy drastically reduces forthcoming infection numbers (decreasing the peak of daily new cases from more than 35,000 to 16,000), hospital and ICU occupancy, and deaths (decreasing the peak of daily deaths from 650 to 280), with a sustained reduction of the epidemic phenomenon over time, which facilitates the vaccination campaign and increases the efficacy and feasibility of testing and tracing[18]. Even though the average $\mathcal{R}_0$ is above 1, the effective reproduction number $\mathcal{R}_t = \mathcal{R}_0 S(t)$ goes below 1 due to the decreasing susceptible fraction $S(t)$, hence the epidemic is eventually suppressed (see Methods).

Finally, we comparatively assess the effect of mass vaccination with different paces. We assume that no reinfections occur in a one-year horizon. **Figure 4** compares the effect of slow vs. fast vaccination under the intermittent Open-Close strategy. Although vaccination leads to a net reduction in deaths and hospital and ICU occupancy compared to the corresponding scenario without vaccination, the difference in impact between slow and fast vaccination is modest: by the time the impact of a faster vaccination could become evident, the epidemic is already contained. The difference is more visible at a higher $\mathcal{R}_0$, at the price of much higher losses. In **Extended Data Figures 6 and 7**, we also consider an adaptive vaccination scenario, where an increase in the number of current infection cases leads to a reduction of the vaccination rate, due to the augmented strain on the healthcare system: both death toll and healthcare system costs increase, reinforcing conclusions about the greater importance of containment measures over vaccination rates.

Our findings confirm that physical distancing, testing and contact tracing are now more crucial than ever due to the circulation of highly transmissible variants of SARS-CoV-2 and the risk of the emergence of vaccine-resistant mutations. We have shown the role of non-pharmaceutical interventions throughout a mass vaccination campaign in keeping the reproduction number low until a sufficient population immunity is achieved. In order to contain the epidemic, restrictions appear more effective than a fast vaccination campaign, in line with US-based studies[29]. Casualties and healthcare system costs predicted by our data-based model also show the importance of pre-emptive action when enforcing intermittent close-open strategies: without any aggravation of socio-economic costs, early closures can drastically lower healthcare system costs with respect to delayed closures of the same duration and stringency.

In our scenarios, vaccination has been assumed effective against SARS-CoV-2 variants. However, several concerns are raised regarding variants and their potential for vaccine-induced immunity escape[30,31]; preliminary reports suggest that COVID-19 vaccines likely retain efficacy against variants[32], although it might be attenuated[33], but data are currently limited. Long-term and large-scale monitoring is required to prove these assumptions; in the meanwhile, enforcing NPIs to keep case numbers low is particularly important.[17,18,31]

**Methods**

Our overall model (see **Figure 1A**) combines the flexibility and insight of compartmental models with the intrinsic robustness of a black-box healthcare system cost model based on observed data. The SIDARTHE-V model, including the compartment of vaccinated individuals (first block in **Figure 1A**), generates the predicted evolution of new positive cases, which feeds the data-based model (second block in **Figure 1A**) that captures hospitalisation flows and quantifies healthcare system costs in terms of deaths and of hospital and ICU occupancy.

*SIDARTHE-V Compartmental Model*

The SIDARTHE-V compartmental model shown in **Figure 1A** extends the SIDARTHE model, first introduced by Giordano et al.[2], by including the effect of vaccination. This leads to nine possible stages of infection: susceptible individuals (S) are uninfected and not immunised; infected individuals (I) are asymptomatic and undetected; diagnosed individuals (D) are asymptomatic but detected; ailing individuals (A) are symptomatic but undetected; recognised individuals (R) are symptomatic and detected; threatened individuals (T) have acute life-threatening symptoms and are detected; healed individuals (H) have had the infection and recovered; extinct individuals (E) died because of the infection; and vaccinated individuals (V) have successfully obtained immunity without having been infected.

The dynamic interaction between these nine clusters of the population is described by the following nine ordinary differential equations, describing how the fraction of the population in each cluster evolves over time:

$$\dot{S}(t) = -S(t)\big(\alpha I(t) + \beta D(t) + \gamma A(t) + \delta R(t)\big) - \varphi(S(t)) \tag{1}$$

$$\dot{I}(t) = S(t)\big(\alpha I(t) + \beta D(t) + \gamma A(t) + \delta R(t)\big) - (\varepsilon + \zeta + \lambda)I(t) \tag{2}$$

$$\dot{D}(t) = \varepsilon I(t) - (\eta + \rho)D(t) \tag{3}$$

$$\dot{A}(t) = \zeta I(t) - (\theta + \mu + \kappa)A(t) \tag{4}$$

$$\dot{R}(t) = \eta D(t) + \theta A(t) - (\nu + \xi + \tau_1)R(t) \tag{5}$$

$$\dot{T}(t) = \mu A(t) + \nu R(t) - (\sigma + \tau_2)T(t) \tag{6}$$

$$\dot{H}(t) = \lambda I(t) + \rho D(t) + \kappa A(t) + \xi R(t) + \sigma T(t) \tag{7}$$

$$\dot{E}(t) = (\tau_1 + \tau_2)T(t) \tag{8}$$

$$\dot{V}(t) = \varphi(S(t)) \tag{9}$$

The uppercase Latin letters (state variables) represent the fraction of population in each stage, while all the considered parameters, denoted by Greek letters, are positive numbers and have the following meaning.

- The *contagion parameters* $\alpha, \beta, \gamma, \delta$ respectively denote the transmission rate (defined as the probability of disease transmission in a single contact multiplied by the average number of contacts per person) due to contacts between a Susceptible subject and an Infected, a Diagnosed, an Ailing, a Recognised subject. These parameters can be modified by social distancing policies (e.g., closing schools, remote working, lockdown), as well as physical distancing, adoption of proper hygiene behaviours and use of personal protective equipment. The risk of contagion due to Threatened subjects, treated in proper ICUs, is assumed negligible.

- The *diagnosis parameters* $\varepsilon$ and $\theta$ respectively denote the probability rate of detection, relative to asymptomatic and symptomatic cases. These parameters, also modifiable, reflect the level of attention on the disease and the number of tests performed over the population: they can be increased by enforcing a massive contact-tracing and testing campaign.

- The *symptom-onset parameters* $\zeta$ and $\eta$ represent the probability rate at which an infected subject, respectively undetected and detected, develops clinically relevant symptoms. Although disease-dependent, they may be partially reduced by improved therapies and acquisition of immunity against the virus.

- The *critical/aggravation parameters* $\mu$ and $\nu$ respectively denote the rate at which undetected and detected infected symptomatic subjects develop life-threatening symptoms. They parameters can be reduced by means of improved therapies and acquisition of immunity against the virus.

- The *mortality parameters* $\tau_1$ and $\tau_2$ respectively denote the mortality rate for infected subjects with symptoms (presumably in hospital wards) and with acute symptoms (presumably in intensive care units) and can be reduced by means of improved therapies.

- The *healing parameters* $\lambda, \kappa, \xi, \rho$ and $\sigma$ denote the rate of recovery for the five classes of infected subjects and can be increased thanks to improved treatments and acquisition of immunity against the virus.

- The *vaccination function* $\varphi(S(t))$ represents the rate at which susceptible people successfully achieve immunity through vaccination (hence it depends on both the actual vaccination rate and the efficacy of the specific vaccine); possible choices are the state-dependent $\varphi(S(t)) = \varphi S(t)$, leading to an exponential decay of the number of susceptible individuals, and $\varphi(S(t)) = \varphi(t) > 0$ as long as $S(t) > 0$ ($\varphi(S(t)) = 0$ otherwise), leading to a linear decay. In the latter case, $\varphi(t)$ can be piecewise-constant, as in the vaccination profiles in **Extended Data Figure 4**.

For an appropriate choice of the vaccination function (such as the two choices proposed above), the bilinear SIDARTHE-V model (1)-(9) is a positive system: all the state variables take nonnegative values for $t \geq 0$ if initialised at time 0 with nonnegative values. Note that $H(t)$, $E(t)$ and $V(t)$ are cumulative variables that depend only on the other ones and on their own initial conditions.

The system is compartmental and has the mass conservation property: as it can be immediately checked, $\dot{S}(t) + \dot{I}(t) + \dot{D}(t) + \dot{A}(t) + \dot{R}(t) + \dot{T}(t) + \dot{H}(t) + \dot{E}(t) + \dot{V}(t) = 0$, hence the sum of the states (total population) is constant. Since the variables denote population *fractions*, we have:

$$S(t) + I(t) + D(t) + A(t) + R(t) + T(t) + H(t) + E(t) + V(t) = 1,$$

where 1 denotes the total population, including deceased.

Given an initial condition $S(0)$, $I(0)$, $D(0)$, $A(0)$, $R(0)$, $T(0)$, $H(0), E(0), V(0)$ summing up to 1, if the vaccination function $\varphi(S(t)) > 0$ as long as $S(t) > 0$, the variables converge to an equilibrium

$$\bar{S} = 0, \bar{I} = 0, \bar{D} = 0, \bar{A} = 0, \bar{R} = 0, \bar{T} = 0, \bar{H} \geq 0, \bar{E} \geq 0, \bar{V} \geq 0,$$

with $\bar{H} + \bar{E} + \bar{V} = 1$. So only the vaccinated/immunised, the healed and the deceased populations are eventually present, meaning that the epidemic phenomenon is over. All the possible equilibria are given by $(0,0,0,0,0,0, \bar{H}, \bar{E}, \bar{V})$, with $\bar{H} + \bar{E} + \bar{V} = 1$.

To better understand the system behaviour, we partition it into three subsystems: the first includes just variable $S$ (corresponding to susceptible individuals); the second, which we denote as the $IDART$ subsystem, includes $I$, $D$, $A$, $R$ and $T$ (the infected individuals); and the third includes variables $H$, $E$ and $V$ (representing healed, defuncts and vaccinated/immunised).

The overall system can be seen as a positive linear system subject to a feedback signal $u$. Defining $x = [I\ D\ A\ R\ T]^\mathsf{T}$, we can rewrite the $IDART$ subsystem as

$$\dot{x}(t) = Fx(t) + bu(t) = \begin{bmatrix} -r_1 & 0 & 0 & 0 & 0 \\ \varepsilon & -r_2 & 0 & 0 & 0 \\ \zeta & 0 & -r_3 & 0 & 0 \\ 0 & \eta & \theta & -r_4 & 0 \\ 0 & 0 & \mu & \nu & -r_5 \end{bmatrix} x(t) + \begin{bmatrix} 1 \\ 0 \\ 0 \\ 0 \\ 0 \end{bmatrix} u(t) \quad (10)$$

$$y_S(t) = c^\mathsf{T} x(t) = [\alpha\ \beta\ \gamma\ \delta\ 0] x(t) \quad (11)$$

$$y_H(t) = f^\mathsf{T} x(t) = [\lambda\ \rho\ \kappa\ \xi\ \sigma] x(t) \quad (12)$$

$$y_E(t) = d^\mathsf{T} x(t) = [0\ 0\ 0\ \tau_1\ \tau_2] x(t) \quad (13)$$

$$u(t) = S(t) y_S(t) \quad (14)$$

where $r_1 = \varepsilon + \zeta + \lambda$, $r_2 = \eta + \rho$, $r_3 = \theta + \mu + \kappa$, $r_4 = \nu + \xi + \tau_1$, $r_5 = \sigma + \tau_2$. The remaining variables satisfy the differential equations

$$\dot{S}(t) = -S(t)y_S(t) - \varphi(S(t)) \tag{15}$$

$$\dot{H}(t) = y_H(t) \tag{16}$$

$$\dot{E}(t) = y_E(t) \tag{17}$$

$$\dot{V}(t) = \varphi(S(t)) \tag{18}$$

We can also distinguish between diagnosed healed $H_D(t)$, evolving as

$$\dot{H}_D(t) = \rho D(t) + \xi R(t) + \sigma T(t),$$

and undiagnosed healed $H_U(t)$, evolving as

$$\dot{H}_U(t) = \lambda I(t) + \kappa A(t).$$

Then, the overall system is described by the infection stage dynamics

$$\dot{x}(t) = (F + bS(t)c^\top)x(t)$$

along with the equations for $S(t)$, $E(t)$ and $H(t)$ (or, equivalently, $H_D(t)$ and $H_U(t)$).

The *parametric reproduction number* $\mathcal{R}_0$ is the $H_\infty$ norm of the positive system from $u$ to $y_S$ with parameters tuned at the beginning of the epidemic, i.e., when the fraction of susceptible individuals is 1. A simple computation leads to

$$\mathcal{R}_0 = \frac{\alpha + \frac{\beta\varepsilon}{r_2} + \frac{\gamma\zeta}{r_3} + \delta(\frac{\eta\varepsilon}{r_2 r_4} + \frac{\zeta\theta}{r_3 r_4})}{r_1}$$

All the parameters but $\varphi$ are represented in $\mathcal{R}_0$. Because these parameters depend on the adopted containment measures, on the effectiveness of therapies and of the efficacy of testing and contact tracing, $\mathcal{R}_0$ is time-varying in principle. Conversely, the *basic reproduction number* is the value of the parametric reproduction number at the first onset of the epidemic outbreak, and its value was estimated to range between 2.43 and 3.1 for SARS-CoV-2 in Italy[35]. The *current reproduction number* $\mathcal{R}_t$ is the product between $\mathcal{R}_0$ and the susceptible fraction: $\mathcal{R}_t = \mathcal{R}_0 S(t)$. Notice that $\mathcal{R}_0$ depends linearly on the contagion parameters. A thorough parameter sensitivity analysis has been worked out for the SIDARTHE model[2].

Fundamental mathematical results on the stability and convergence of the model are summarised next.

*The system (10), (14) with constant parameters and constant susceptible population $S(\bar{t})$ is asymptotically stable if and only if $\mathcal{R}_{\bar{t}} < 1$. The equilibrium point $\bar{x}, \bar{S}$ of system (10), (14) with constant parameters after $\bar{t}$ is given by $\bar{x} = 0$ and $\bar{S}$ satisfying*

$$\ln\left(\frac{S(\bar{t})}{\bar{S}}\right) + \mathcal{R}_0(\bar{S} - S(\bar{t})) = -c^\top F^{-1} x(\bar{t})$$

*The condition $\mathcal{R}_t = \mathcal{R}_0 S(t) < 1$ is always verified after a certain time instant, so that $x(t) \to 0$ and $S(t) \to \bar{S}$ with $\mathcal{R}_0 \bar{S} < 1$.*

As a consequence, epidemic suppression is achieved when the inequality $\mathcal{R}_t = \mathcal{R}_0 S(t) < 1$ is always verified from a certain moment onwards.

### *Fit of the SIDARTHE-V Model for the COVID-19 Epidemic in Italy*

We infer the parameters for model (1)-(9) based on the official data (source: Protezione Civile and Ministero della Salute) about the evolution of the epidemic in Italy from February 24th, 2020 (day 1) through February 4th, 2021 (day 346). We turn the data into fractions over the whole Italian population (~60 million) and adopt a best-fit approach to find the parameters that locally minimize the sum of the squares of the errors.

With parameters estimated based on data till February 4th, 2021, the SIDARTHE-V model reproduces the second wave of infection and feeds the health cost model that quantifies the healthcare system costs in different scenarios, encompassing close-open strategies during the mass vaccination campaign.

The validation in **Extended Data Figure 1** shows how the SIDARTHE model (initially without vaccination) can faithfully reproduce the epidemic evolution observed so far. In the figures, the evolution over time of the number of active cases, hospitalisations and ICU occupancy, as well as daily deaths, is reported in logarithmic scale, comparing data (dots) with the model prediction (solid line). After March 8th, 2020, a strict lockdown brought the Italian effective reproduction number below 1, successfully reversing the COVID-19 epidemic trend. Commercial and recreational activities gradually reopened, until the lockdown was fully lifted on June 3rd. The lockdown relaxation coincided with a decreased risk perception and increased social gatherings when, after the first wave, restrictions were eased during the summer. Hence, as expected, a new upward trend in SARS-CoV-2 infections began in mid-August. The increase in daily cases, slow and steady at first, eventually led to a failure in the contact tracing system and the occurrence of a second wave. School reopening in the third week of September led to a steady increase in the number of new cases, hospital and ICU occupancy. This prompted the Italian government, on November 4th, to introduce a three-tier system enforcing diversely strict containment measures on a regional basis, depending on different risk scenarios. The slow decrease of the reproduction and hospitalization numbers led to stricter rules for the period December 24th-January 6th. The initial onset of the second wave can be promptly identified at the beginning of August 2020, when the infection variables reach a minimum value. The descent phase of the second wave was much slower than that of the first, revealing that the enforced containment measures were milder. In particular, progressive countermeasures were enforced on October 24th (partial limitations), November 4th (regional lockdowns) and December (country-wide lockdown). Mild easing of restrictions, and school reopening, started on January 7th, 2021, while other regional measures were implemented on January 15th and subsequently eased.

To reproduce the epidemic evolution over time, the system parameters are piecewise constant and are possibly updated at the following days:

[1 4 12 22 28 36 38 40 47 60 75 119 151 163 182 213 221 253 258 275 293 308 320 325 328 346]

The chosen parameter values are

$\alpha$= [0.6588 0.4874 0.4886 0.3467 0.2311 0.2543 0.2543 0.3174 0.3467 0.3351 0.3236 0.3236 0.3467 0.4391 0.4045 0.3699 0.4869 0.4166 0.3567 0.3567 0.3267 0.4531 0.3234 0.4431 0.4132 0.3134]

$\beta$=[ 0.012 0.006 0.006 0.0053 0.0053 0.0053 0.0053 0.0053 0.0053 0.0053 0.0053 0.0053 0.0053 0.0053 0.0053 0.0053 0.0053 0.0053 0.0053 0.0053 0.0053 0.0053 0.0053 0.0053 0.0053 0.0053]

$\gamma$=[0.4514 0.2821 0.2821 0.1980 0.1089 0.1089 0.1089 0.1089 0.1089 0.1188 0.1188 0.1485 0.1485 0.1485 0.1485 0.1485 0.1485 0.1485 0.1485 0.1485 0.1485 0.1485 0.1485 0.1485 0.1485 0.1485]

δ=[0.0113 0.0056 0.0056 0.005 0.005 0.005 0.005 0.005 0.005 0.005 0.005 0.005 0.005 0.005 0.005 0.005 0.005 0.005 0.005 0.005 0.005 0.005 0.005 0.005 0.005 0.005]

ε=[0.1703 0.1703 0.1419 0.1419 0.1419 0.1419 0.1992 0.2988 0.2988 0.2988 0.2988 0.6972 0.2490 0.2988 0.2590 0.2294 0.3137 0.2868 0.2868 0.2988 0.2988 0.2988 0.2988 0.3988 0.3988 0.1988]

θ=[0.3705 0.3705 0.3705 0.3705 0.3705 0.3705 0.3705 0.5 0.5 0.5 0.5 0.5 0.5 0.6 0.3 0.6 0.37 0.370 0.37 0.37 0.37 0.37 0.37 0.37 0.37 0.37]

ζ=[0.1254 0.1254 0.1254 0.0340 0.0340 0.0341 0.0250 0.0250 0.0015 0 0.0001 0.0001 0.0005 0.0020 0.0030 0.0020 0.0046 0.0025 0.0025 0.0025 0.0025 0.0025 0.0025 0.0025 0.0025 0.0025]

ε=[0.1054 0.1054 0.1054 0.0286 0.0286 0.0286 0.0286 0.021 0.0015 0 0 0 0.0005 0.002 0.0031 0.0026 0.003 0.0013 0.0013 0.001 0.0015 0.0018 0.0018 0.0018 0.0018 0.0018]

μ=[0.0205 0.0205 0.0205 0.0096 0.0084 0.0036 0.0036 0.0036 0 0 0 0.0036 0.0036 0 0.0024 0.0036 0.06 0.12 0.12 0. 0.12 0.12 0.12 0.12 0.12 0.12 0.12]

ν=[0.03 0.03 0.01 0.01 0.01 0.008 0.007 0.006 0.005 0.004 0.025 0.025 0.0026 0.0026 0.0026 0.002 0.002 0.002 0.002 0.02 0.02 0.02 0.02 0.02 0.02 0.02]

τ2=[0 0 0 0 0 0 0.035 0.045 0.045 0.045 0.4500 0.45 0.02 0 0 0 0 0.0005 0.0005 0.17 0.17 0.17 0.17 0.17 0.17 0.17]

τ1=[0.02 0.02 0.02 0.02 0.02 0.02 0.02 0.05 0.01 0.01 0 0 0 0 0 0 0.018 0.018 0.001 0.001 0.005 0.001 0.005 0.005 0.005 0.005]

λ=[0.0482 0.0482 0.0482 0.1128 0.1128 0.1128 0.1128 0.1128 0.1128 0.1128 0.1128 0.1128 0.1128 0.1128 0.1128 0.1128 0.1128 0.1128 0.1128 0.1128 0.1128 0.1128 0.1128 0.1128 0.1128 0.1128]

ρ=[0.0342 0.0342 0.0342 0.017 0.017 0.017 0.02 0.022 0.022 0.045 0.045 0.045 0.02 0.018 0.018 0.018 0.018 0.018 0.032 0.032 0.032 0.032 0.032 0.032 0.032 0.032]

κ=[0.0171 0.0171 0.0171 0.0171 0.0171 0.0171 0.02 0.022 0.022 0.035 0.035 0.035 0.02 0.02 0.02 0.02 0.02 0.02 0.02 0.02 0.02 0.02 0.02 0.02 0.02 0.02]

χ=[0.00025 0.00025 0.00025 0.00025 0.00025 0.00025 0.00025 0.0083 0.0083 0.0207 0.012 0.012 0.0037 0.0019 0.0019 0.00067 0.00067 0.00067 0.00015 0.022 0.022 0.032 0.022 0.0220 0.022 0.022]

σ=[0.0513 0.0513 0.0513 0.0513 0.0513 0.0513 0.03 0.03 0.06 0.075 0.003 0 0 0 0.024 0.024 0.024 0.024 0.0024 0.024 0.024 0.024 0.024 0.024 0.024 0.024]

and lead to the following piecewise constant parametric reproduction number:

$\mathcal{R}_0$=[2.5200  1.7692  1.9725  1.3859  0.9593  1.0448  0.9162  0.8533  1.0138  0.8997  0.8715  0.5008  1.1494  1.3398  1.3776  1.4154  1.3439  1.2464  1.0104  0.9827  0.9097  0.9089  1.2131  0.9003  0.7374  0.9897  0.8779 ]

Starting with day 255, the parameters are differentiated depending on the different scenarios associated with the presence of new virus variants and/or of the adoption of different restrictions:

- High transmission: $\alpha$ =0.477, leading to a constant $\mathcal{R}_0 = 1.27$
- Open-Close: $\alpha$ switches every month between the high value 0.477 and the low value 0.3198, hence every month $\mathcal{R}_0$ switches between the values 1.3 and 0.9, with an average value of 1.1.
- Constant $\alpha$ =0.4092, leading to $\mathcal{R}_0 = 1.1$
- Close-Open: $\alpha$ switches every month between the low value 0.3198 and the high value 0.477, hence every month $\mathcal{R}_0$ switches between the values 0.9 and 1.3, with an average value of 1.1. The Close-Open strategy features the same pattern as the Open-Close strategy, the only difference being that is starts with a closure phase.
- Eradication: $\alpha$ =0.3198, leading to a constant $\mathcal{R}_0 = 0.9$.

The vaccination function is chosen according to the three different profiles show in **Extended Data Figure 4**, where $\varphi(S(t)) = \varphi(t)$ is piecewise constant (of course, $\varphi(S(t)) = 0$ when $S(t) = 0$). In the adaptive vaccination scenarios in **Extended Data Figures 6 and 7**, conversely, the vaccination function is chosen as $\max\{[1 - r(D(t) + R(t) + T(t))]\varphi(t); 0\}$, where $\varphi(t)$ is the same piecewise constant vaccination profile as above, while the parameter $r$ is chosen as $r = 10^{-6}$.

Adaptive vaccination functions with slow/medium/fast vaccination schedules (see **Extended Data Figures 6 and 7**) lead to increased healthcare system costs with respect to piecewise-constant vaccination functions (**Figure 4**), for all $\mathcal{R}_0$ profiles. When $\mathcal{R}_0 = 1.27$, thousand deaths in the period from February 2021 to January 2022 increase from 196 to 330 (slow speed), from 174 to 325 (medium speed), from 146 to 319 (fast speed), while they would be 352 without vaccination. Under the Open-Close strategy with average $\mathcal{R}_0$ equal to 1.1, thousand deaths in the period from February 2021 to January 2022 increase from 66 to 99 (slow speed), from 60 to 94 (medium speed), from 54 to 87 (fast speed), while they would be 130 without vaccination. When $\mathcal{R}_0 = 1.1$, thousand deaths in the period from February 2021 to January 2022 increase from 47 to 69 (slow speed), from 43 to 65 (medium speed), from 38 to 59 (fast speed), while they would be 109 without vaccination. Under the Close-Open strategy with average $\mathcal{R}_0$ equal to 1.1, thousand deaths in the period from February 2021 to January 2022 increase from 39 to 57 (slow speed), from 35 to 52 (medium speed), from 31 to 47 (fast speed), while they would be 100 without vaccination.

### *Data-Driven Model of Healthcare System Costs*

In order to predict the evolution of deaths from the time series of reported cases, a field estimate of the apparent CFR (Case Fatality Rate) is needed. This parameter is affected by the testing protocol, the healthcare system reaction, and the age distribution of vaccinated people. For these reasons, a specific model should be derived for each country resorting to a data-based approach. We considered the Italian case, but the methodology has general validity and could be promptly applied to other countries. The input-output model predicting deaths from the new cases was derived in two steps. First, a data-based dynamic model for the unvaccinated population was estimated from data collected during the second wave. The static gain of this dynamic model coincides with the CFR for the unvaccinated population. In the second step, this gain was multiplied by a time varying function that accounts for the lethality decrease consequent to progressive vaccination of older people.

The input of the dynamic model is the time series of new cases $n(t)$ and the output is the time series of daily deaths $d(t)$. Given, the variations of testing protocols and the large number of unreported case during the first wave, the model was estimated using data collected within a 110-day window ending on February 7, 2021. The model assumes that the deaths at day $t$ depend on the past new cases according to the equation

$$d(t) = \sum_{i=0}^{\infty} w(i) n(t-i)$$

where the weight $w(i)$ denotes the fraction of subjects that became infected at day $t-i$ that eventually die at day $t$. The CFR for the unvaccinated population is then given by

$$CFR_0 = \sum_{i=0}^{\infty} w(i)$$

In order to estimate the weights, an exponential model with delay was assumed:

$$w(i) = \begin{cases} 0, & i < k \\ bf^{i-k}, & i \geq k \end{cases}$$

where $k$ is the delay and $f$ and $b$ are unknown parameters. Since both the new cases and the daily deaths exhibit an apparent weekly seasonality, the original series were replaced by their 7-day moving average. Parameters were estimated via least squares using the function oe.m of the MATLAB System Identification Toolbox2[36]. The estimation of $a$ and $b$ was repeated for all delays $k$ ranging from 0 to 15 and the delay $k = 3$, associated with the best sum of squares was eventually selected. The estimated parameters and their percent coefficients of variation were:

$$f = 0.948, CV(\%) = 0.311$$

$$b = 0.00141, CV(\%) = 5.36$$

Then, recalling the formula for the sum of the harmonic series, the second-wave CFR for the unvaccinated population is given by

$$CFR_0 = \sum_{i=0}^{\infty} w(i) = \frac{b}{1-f} = 0.0272$$

In the second step, the effect of vaccination on lethality was modelled by estimating a time-varying $CFR(t)$ that depends on the vaccination schedule, described by the fraction $V(t)$ of vaccine-immunized subjects at time $t$. Order of vaccination follows the reverse of the age. Slower or faster vaccination speeds correspond to different curves $V(t)$, whose rate of increase may be less or more rapid. Three schedules were considered: fast, medium, and slow. The fast schedule assumes that each the four phases T1-T4 of the Italian vaccination plan[37] is completed in one trimester. In the medium and slow schedule the time was linearly extended by a factor 1.2 and 1.4, respectively. The three schedules are graphically displayed in **Extended Data Figure 4**.

The COVID-19 lethality $CFR_a$ for a subject of age $a$ was obtained by rescaling values published the Italian National Institute of Health[38]. Rescaling was necessary because the published values were inflated by the inclusion of patients died during the first wave, when new cases were massively underreported. Rescaling was performed in such a way that the overall lethality coincides with $CFR_0 = 0.0272$. The profile $CFR_a$ is displayed in **Extended Data Figure 5**, Panel C.

If we assume that the number of deaths does not affect significantly the overall age distribution, the probability $P(Age = a)$ can be directly inferred by ISTAT statistical tables[39]. The distribution of population by age is displayed in **Extended Data Figure 5**, Panel C.

Recalling that $V(t)$ is the fraction of vaccine-immunized subjects and vaccination order follows the reverse of the age, the probability of death for a subject of age a that becomes infected at time t is

$$P(Death|Age = a, Infected, t) = \begin{cases} 0, & V(t) > P(Age > a) \\ CFR(a), & V(t) \leq P(Age > a) \end{cases}$$

Then, the time varying case fatality rate is obtained by the total probability theorem:

$$CFR(t) = \sum_{a=0}^{100} P(Death|Age = a, Infected, t) P(Age = a|Infected, t)$$

where $P(Age = a|Infected, t)$ denotes the probability that the age of a subject is $a$, knowing that it became infected at time $t$. Currently, the age distribution of the infected subjects[38] appears similar to the age distribution of Italian population. For instance, 56% of population is in the age range 0-50, 28% is in the range 51-70 and 16% is older than 70. In comparison, 55,6% of diagnosed cases between 18/12/2020 and 10/01/2021 are in age range 0-50, 28% in the range 51-70 and 16.4% over 70. Therefore, $P(Age = a|Infected, t) = P(Age = a)$, so that the CFR of a subject infected at time $t$ can be computed as

$$CFR(t) = \sum_{a=0}^{100} P(Death|Age = a, Infected, t) P(Age = a)$$

The steps of the procedure for the computation of $CFR(t)$ are summarized in **Extended Data Figure 5**. The time varying profiles $CFR(t)$ for the three vaccination schedules are plotted in **Extended Data Figure 4**.

Finally, the input-output model that accounts for the effect of vaccination is given by

$$d(t) = \sum_{i=0}^{\infty} w(i) C(t-i) n(t-i), \qquad C(t) = \frac{CFR(t)}{CFR_0}$$

Due to the time varying coefficient $C(t)$, the same number of new cases will yield fewer and fewer deaths as vaccination comes to protect older segments of the population.

Dynamic models for hospital and ICU occupancies were developed in a similar way. The estimated parameters and their percent coefficients of variation for hospital occupancy (estimated delay $k = 0$) were:

$$a = 0.9522, CV(\%) = 0.226$$

$$f = 0.0694, CV(\%) = 4.26$$

Those for ICU occupancy (estimated delay $k = 0$) were:

$$a = 0.953, CV(\%) = 0.328$$

$$f = 0.00677, CV(\%) = 6.297$$

Assuming that gravity reduction parallels the lethality one, the effect of vaccination on hospital and ICU occupancies was described by modulating the input through the time-varying coefficient $C(t)$. As seen in **Extended Data Figure 9**, the three data-based dynamic models provide a very good fitting of deaths, hospital and ICU occupancies.

**Reporting Summary.** Further information on research design is available in the Life Sciences Reporting Summary linked to this article.

**Data availability**

We gathered all epidemiological data from publicly available sources:

https://github.com/pcm-dpc/COVID-19/tree/master/dati-andamento-nazionale.

Data are also included in the Extended Data Figures 1 and 9.

**Code availability**

The codes will be made available online.

**Acknowledgements**

We acknowledge financial support through the Italian grant PRIN 2017 "Monitoring and Control Underpinning the Energy-Aware Factory of the Future: Novel Methodologies and Industrial Validation", ID 2017YKXYXJ. This research has also received funding from the European Union's Horizon 2020 research and innovation program "PERISCOPE: Pan European Response to the ImpactS of COvid-19 and future Pandemics and Epidemics", under the grant agreement No. 101016233, H2020-SC1-PHE-CORONAVIRUS-2020-2-RTD.

**Author contributions**

F.B., P.B., P.C., G.D.N. and G.G. proposed the model and performed the model fitting, simulations, analysis and interpretation of the results. R.B., M.C., A.D.F. and P.S. provided first-hand clinical insight and contextualisation. All authors wrote and approved the manuscript.

**Competing interests**

The authors declare no competing interests.

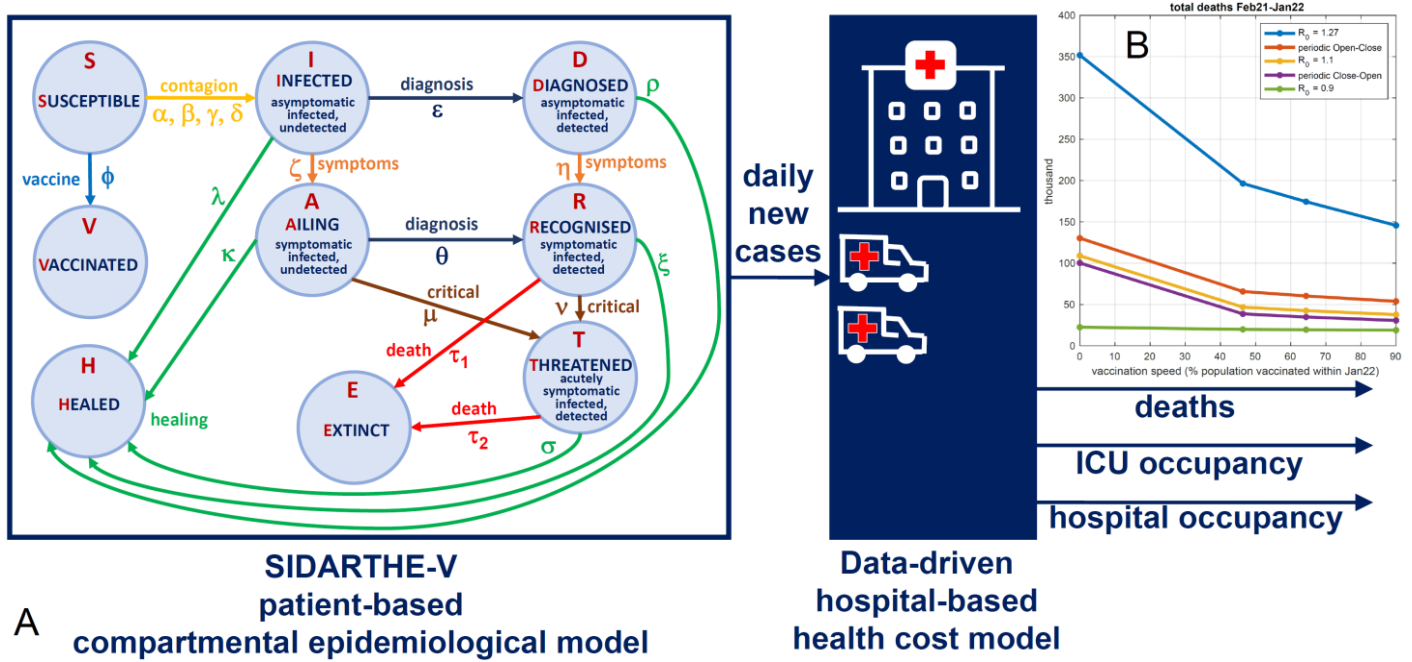

**Figure 1. Model scheme and summary of our findings. A:** Graphical scheme of our model. The compartmental SIDARTHE-V epidemiological model feeds a novel data-based model of casualties and healthcare system costs. The SIDARTHE-V model captures the dynamic interactions among nine mutually exclusive infection stages in the population: **S**, Susceptible (uninfected); **I**, Infected (asymptomatic infected, undetected); **D**, Diagnosed (asymptomatic infected, detected); **A**, Ailing (symptomatic infected, undetected); **R**, Recognised (symptomatic infected, detected); **T**, Threatened (infected with life-threatening symptoms, detected); **H**, Healed (recovered); **E**, Extinct (dead); **V**, Vaccinated (successfully immunised). The SIDARTHE-V model provides the time evolution of daily new infection cases, based on which the data-driven model computes the evolution of deaths, ICU and hospital occupancy. **B:** Our main findings summarised by vaccination-cost curves: for a given $\mathcal{R}_0$ profile, the curve gives the death toll in the period from February 2021 to January 2022 as a function of the average vaccination speed, measured as the fraction of vaccinated population at the end of the period. The vaccination-cost curves corresponding to a constant reproduction number are reported in blue ($\mathcal{R}_0 = 1.27$), yellow ($\mathcal{R}_0 = 1.1$) and green ($\mathcal{R}_0 = 0.9$), while those corresponding to intermittent strategies are reported in orange (Open-Close) and purple (Close-Open).

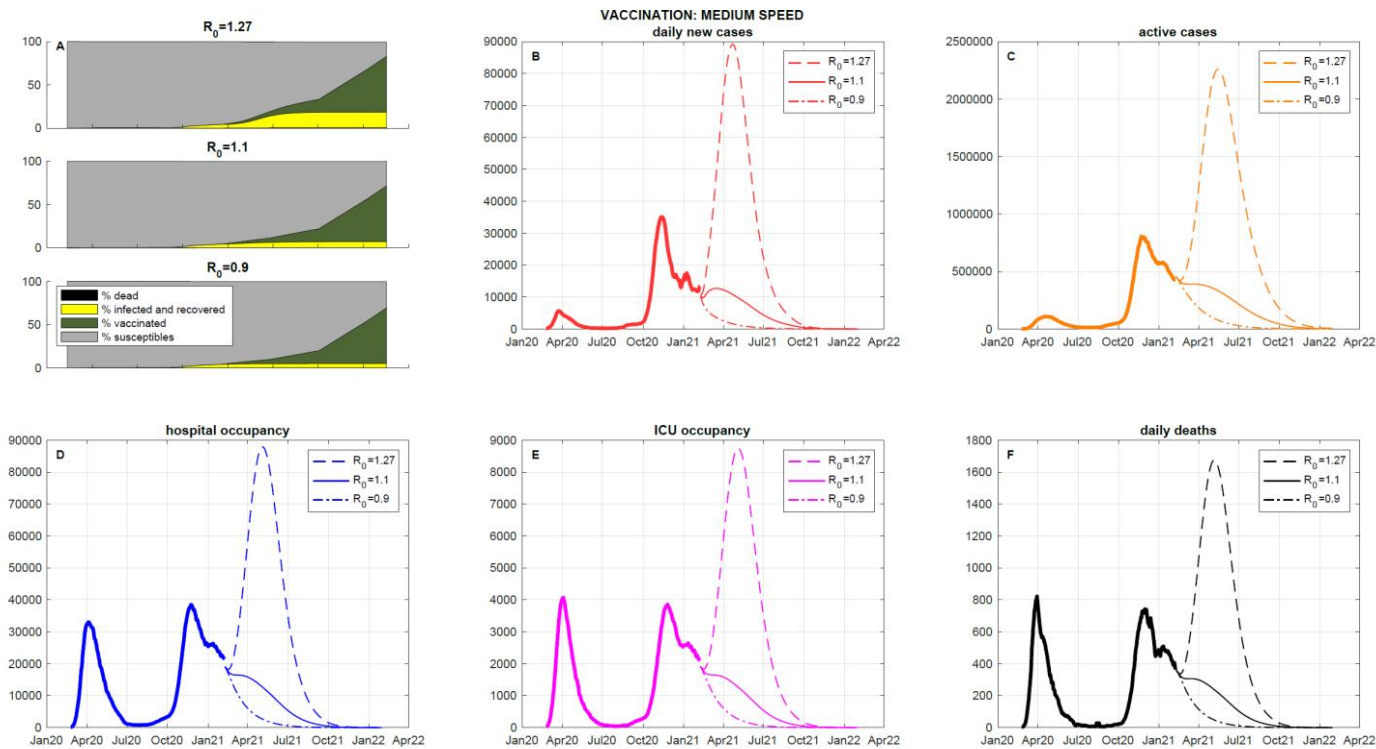

**Figure 2. The impact of different constant values of $\mathcal{R}_0$.** Time evolution of the epidemic, in the presence of medium-speed vaccination (64% of the population vaccinated within one year), when different constant values of $\mathcal{R}_0$, namely $\mathcal{R}_0 = 1.27$ (dashed line), $\mathcal{R}_0 = 1.1$ (solid line), $\mathcal{R}_0 = 0.9$ (dash-dotted line), are assumed, resulting from different variants and/or containment strategies. **A:** Time evolution of the fractions of susceptibles, vaccinated, infected and recovered, and dead. **B:** Daily new cases. **C:** Active cases. **D:** Hospital occupancy. **E:** ICU occupancy. **F:** Daily deaths.

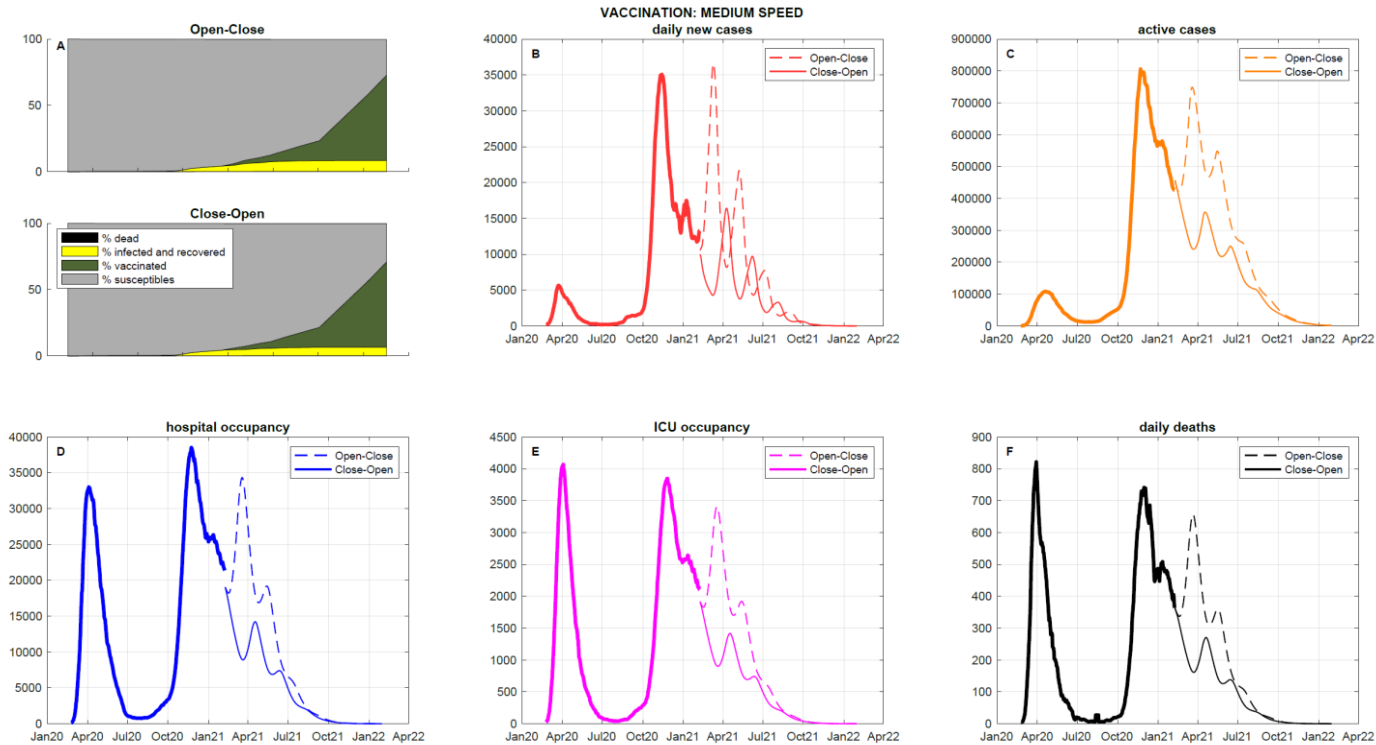

**Figure 3. The impact of different intermittent strategies.** Time evolution of the epidemic, in the presence of medium-speed vaccination (64% of the population vaccinated within one year), when two alternative intermittent strategies are enforced, with an average value of $\mathcal{R}_0$ equal to 1.1. The Open-Close strategy (dashed line) switches every month between $\mathcal{R}_0 = 1.27$ and $\mathcal{R}_0 = 0.9$, starting with $\mathcal{R}_0 = 1.27$. The Close-Open strategy (solid line) switches every month between $\mathcal{R}_0 = 0.9$ and $\mathcal{R}_0 = 1.27$, but starting with $\mathcal{R}_0 = 0.9$. **A:** Time evolution of the fractions of susceptibles, vaccinated, infected and recovered, and dead. **B:** Daily new cases. **C:** Active cases. **D:** Hospital occupancy. **E:** ICU occupancy. **F:** Daily deaths.

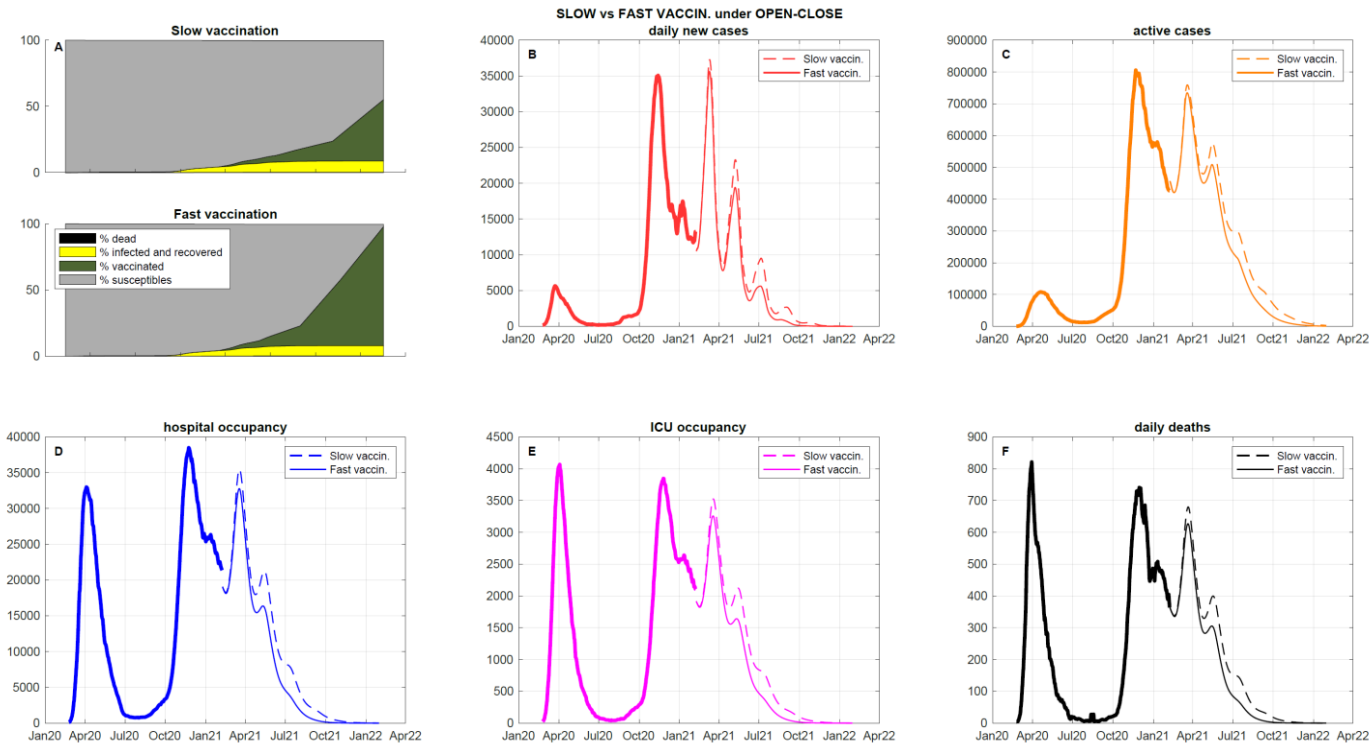

**Figure 4. The impact of different vaccination paces.** Time evolution of the epidemic, with an intermittent Open-Close strategy enforced, in the presence of slow vaccination (46% of the population vaccinated within one year, dashed line) or fast vaccination (90% of the population vaccinated within one year, solid line). **A:** Time evolution of the fractions of susceptibles, vaccinated, infected and recovered, and dead. **B:** Daily new cases. **C:** Active cases. **D:** Hospital occupancy. **E:** ICU occupancy. **F:** Daily deaths.

**Table 1: Policy Summary**

| Background | The second wave of the SARS-CoV-2 pandemic has severely affected Italy, with a high case fatality rate. Two potential game-changers now affect the evolution of the epidemic: the availability of vaccines and the emergence of more transmissible, and perhaps more lethal, virus variants. We combine our compartmental epidemiological model with a new data-based model of healthcare costs to assess the impact of the vaccination campaign on the future evolution of the epidemic, in the presence of different non-pharmaceutical interventions and different SARS-CoV-2 variants. |
|---|---|
| **Main findings and limitations** | Even though mass vaccination has started, non-pharmaceutical interventions remain crucial to control the epidemic, also due to the circulation of highly transmissible variants of SARS-CoV-2. Restrictions curb transmission much more than a fast vaccination: easing them unavoidably leads to a surge of infection cases, calling for new closures, thus triggering intermittent strategies. Pre-emptive strategies (first close, then open at low case numbers) drastically reduce hospitalisations and deaths with respect to a delayed intervention (first open, then close to prevent ICU saturation) without aggravating socioeconomic costs. Our scenarios are outlined based on reasonable assumptions, but the actual epidemic evolution will depend on the adopted measures and the possible emergence of other variants. |
| **Policy implications** | Our findings strongly advocate for the need to keep non-pharmaceutical interventions in place throughout the vaccination campaign, until sufficient population immunity is reached. We also show the great effectiveness of pre-emptive action: the mere anticipation of the closing phase within an intermittent Open-Close strategy spares tens of thousands of deaths and huge healthcare costs, without aggravating socioeconomic losses. |

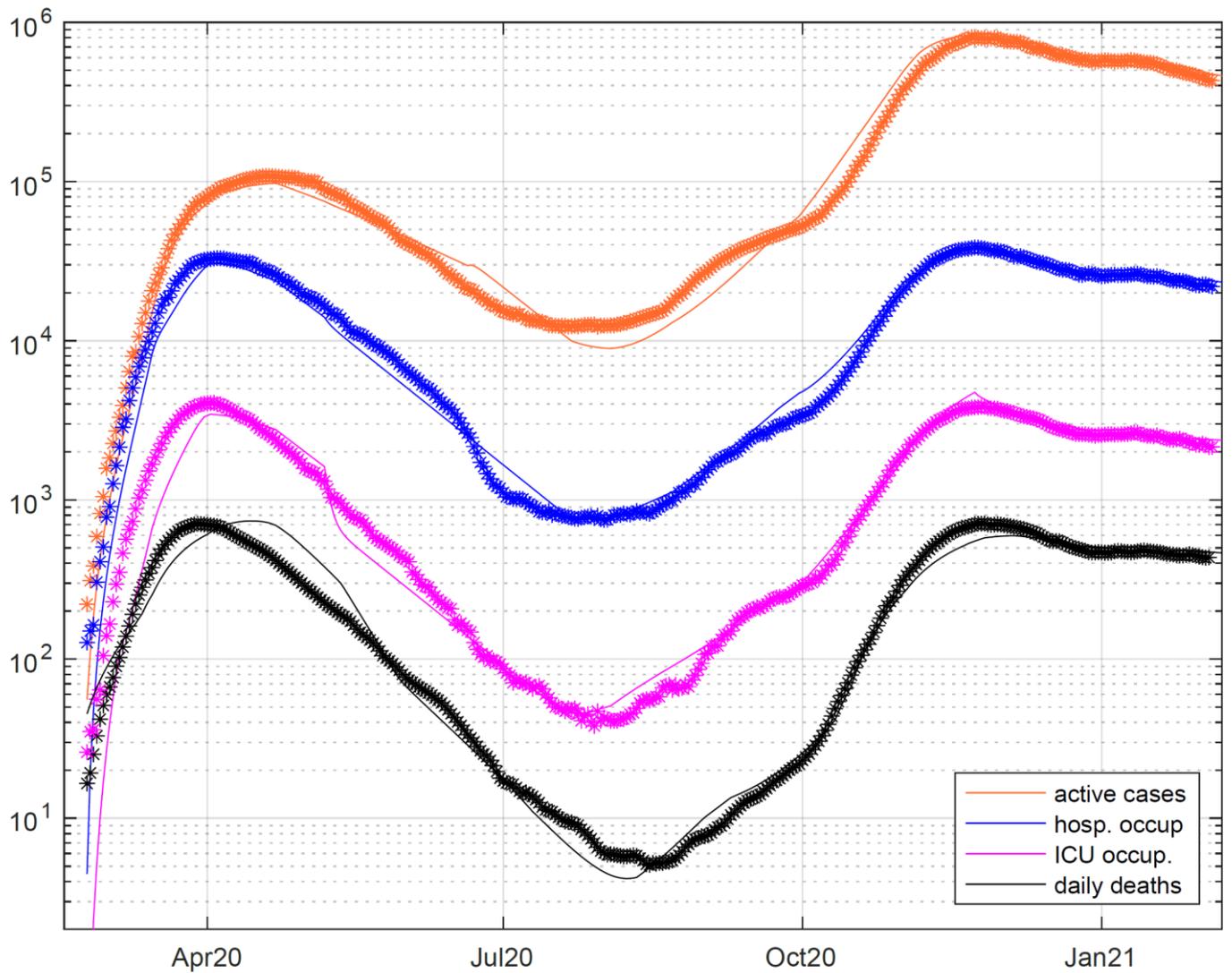

**Extended Data Figure 1.** Epidemic evolution in Italy from late February 2020 to early February 2021: data (stars) and estimation (solid lines), based on the SIDARTHE model, of the time evolution of the epidemic, in logarithmic scale. Active cases (current diagnosed infected, related to the SIDARTHE variables R+T+D) are shown in orange; hospital occupancy (related to the SIDARTHE variables R+T) is shown in blue; ICU occupancy (related to variable T) is shown in magenta; daily deaths (related to the derivative of variable E) are shown in black.

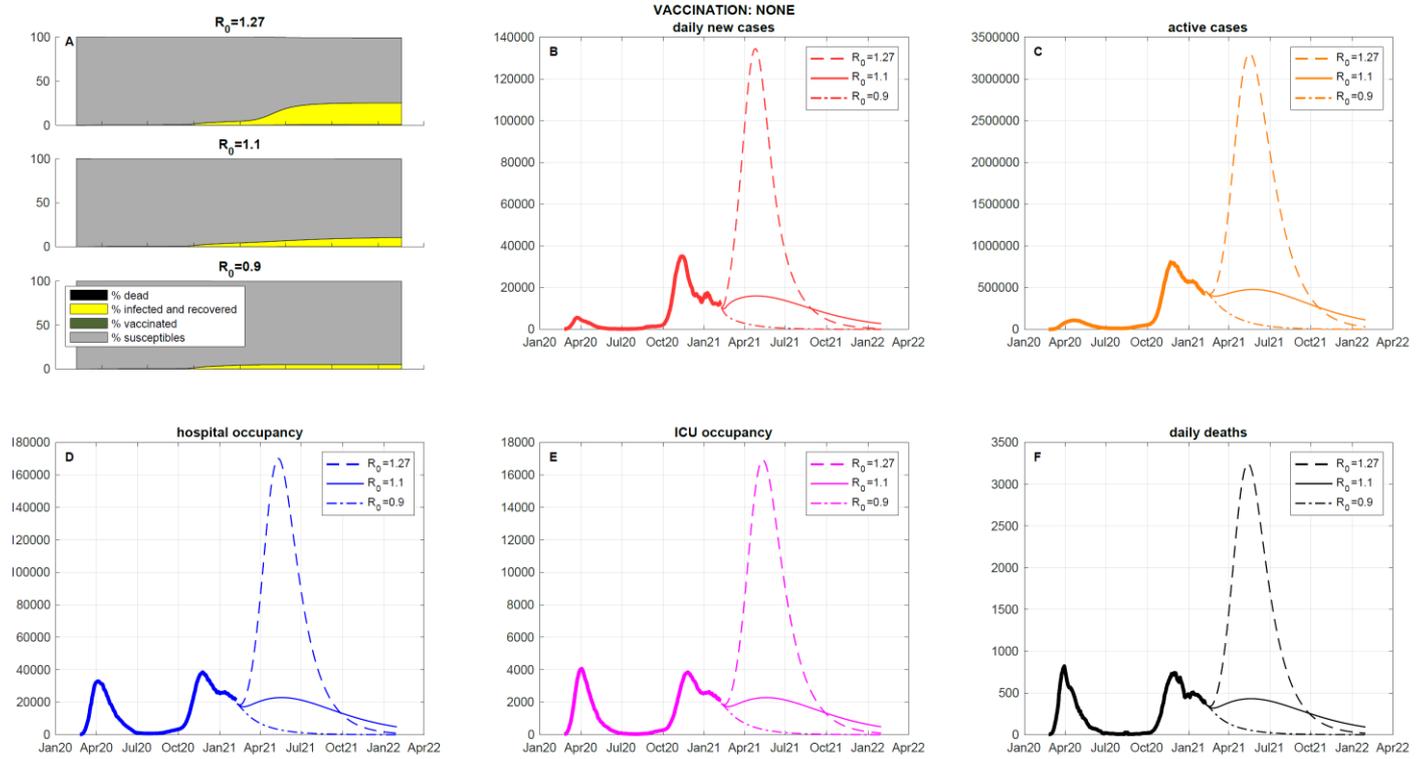

**Extended Data Figure 2.** Time evolution of the epidemic, in the absence of vaccination, when different constant values of $\mathcal{R}_0$, namely $\mathcal{R}_0 = 1.27$ (dashed line), $\mathcal{R}_0 = 1.1$ (solid line), $\mathcal{R}_0 = 0.9$ (dash-dotted line), are assumed, resulting from different variants and/or containment strategies. **A:** Time evolution of the fractions of susceptibles, infected and recovered, and dead. **B:** Daily new cases. **C:** Active cases. **D:** Hospital occupancy. **E:** ICU occupancy. **F:** Daily deaths.

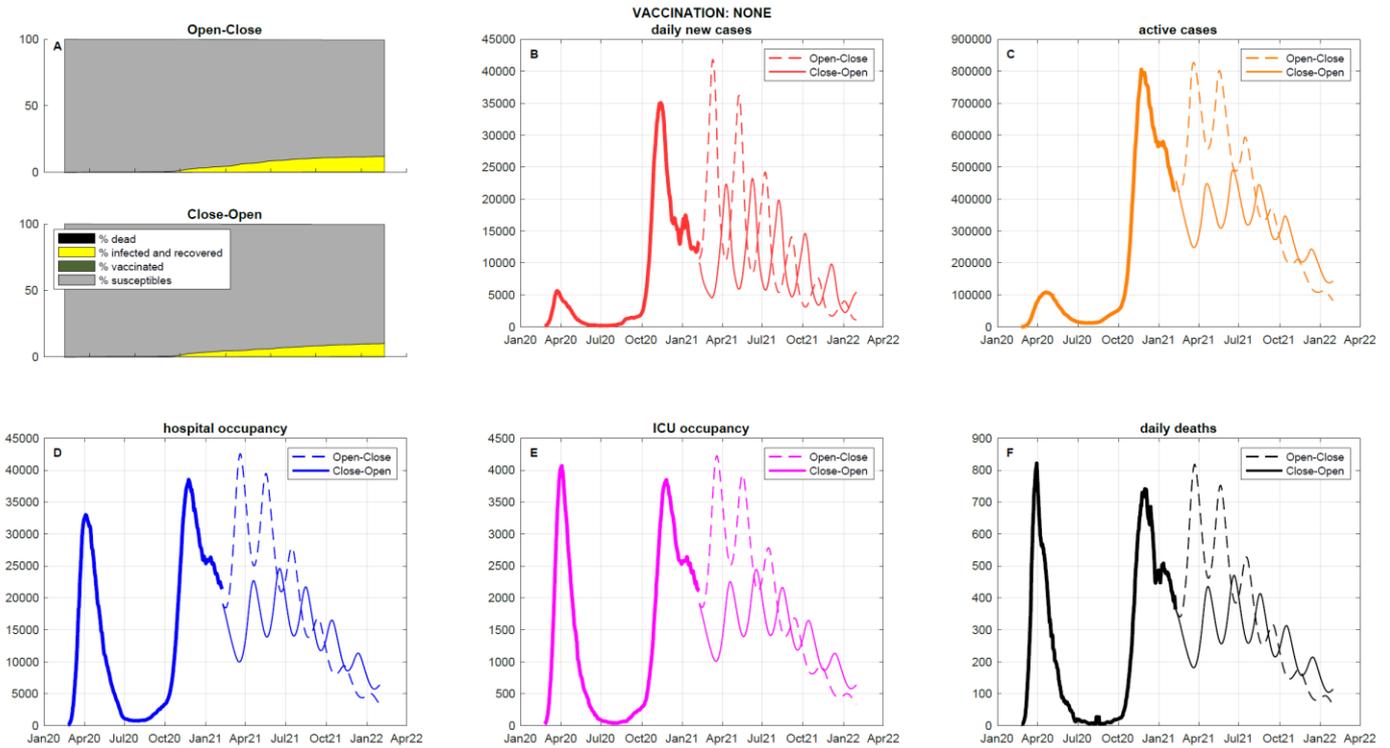

**Extended Data Figure 3.** Time evolution of the epidemic, in the absence of vaccination, when two alternative intermittent strategies are enforced, with an average value of $\mathcal{R}_0$ equal to 1.1. The Open-Close strategy (dashed line) switches every month between $\mathcal{R}_0 = 1.27$ and $\mathcal{R}_0 = 0.9$, starting with $\mathcal{R}_0 = 1.27$. The Close-Open strategy (solid line) does the same, but starts with $\mathcal{R}_0 = 0.9$. **A:** Time evolution of the fractions of susceptibles, infected and recovered, and dead. **B:** Daily new cases. **C:** Active cases. **D:** Hospital occupancy. **E:** ICU occupancy. **F:** Daily deaths.

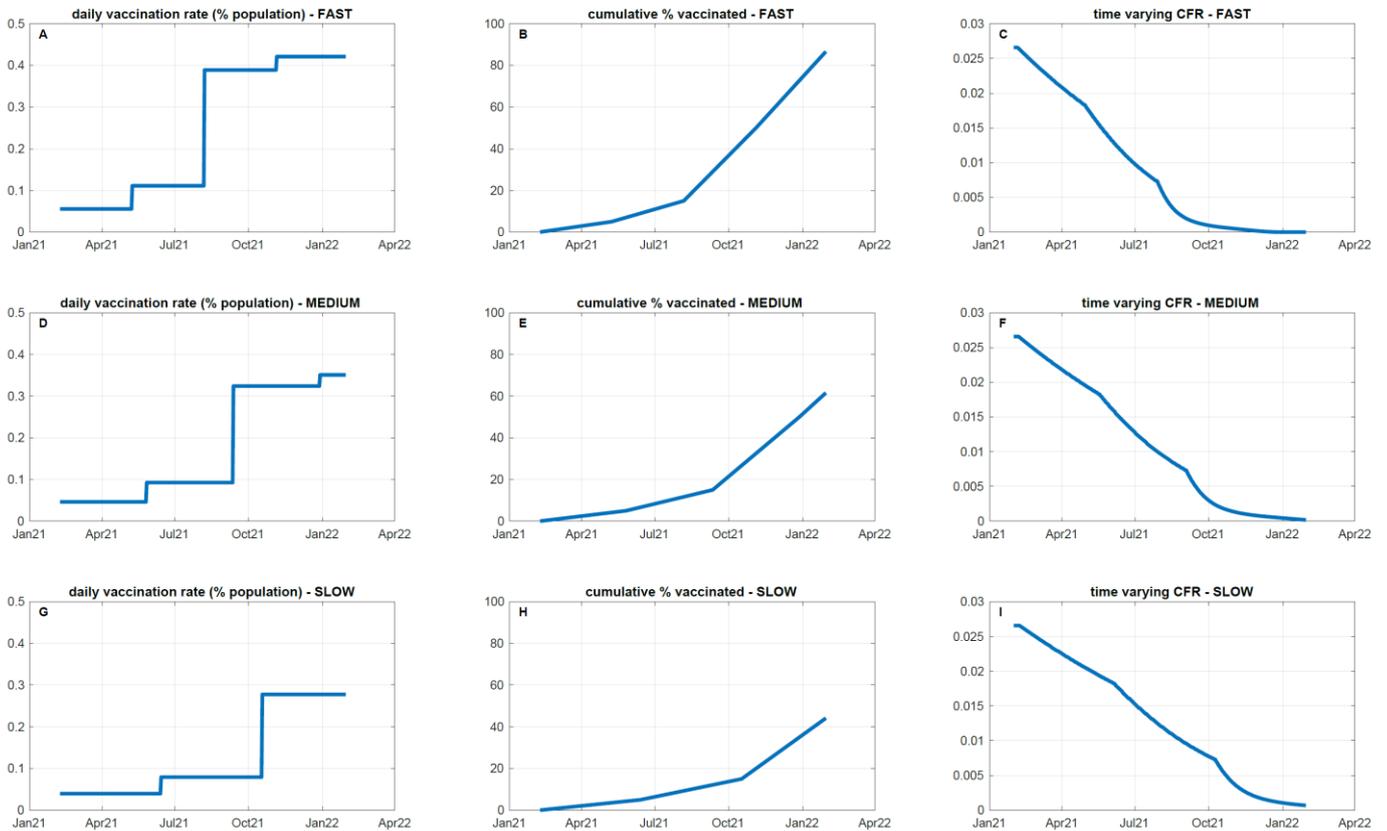

**Extended Data Figure 4.** Profiles of the considered three different effective vaccination schedules: slow, medium and fast. For each of the three schedules, we show: the evolution over time of the daily effective vaccination rate, namely the fraction of population successfully immunised in one day (**A-D-G**), the cumulative fraction of immunised population as a function of time (**B-E-H**), the resulting time-varying Case Fatality Rate as a function of time, obtained taking population age classes into account and assuming a vaccination schedule that prioritises the elderly (**C-F-I**).

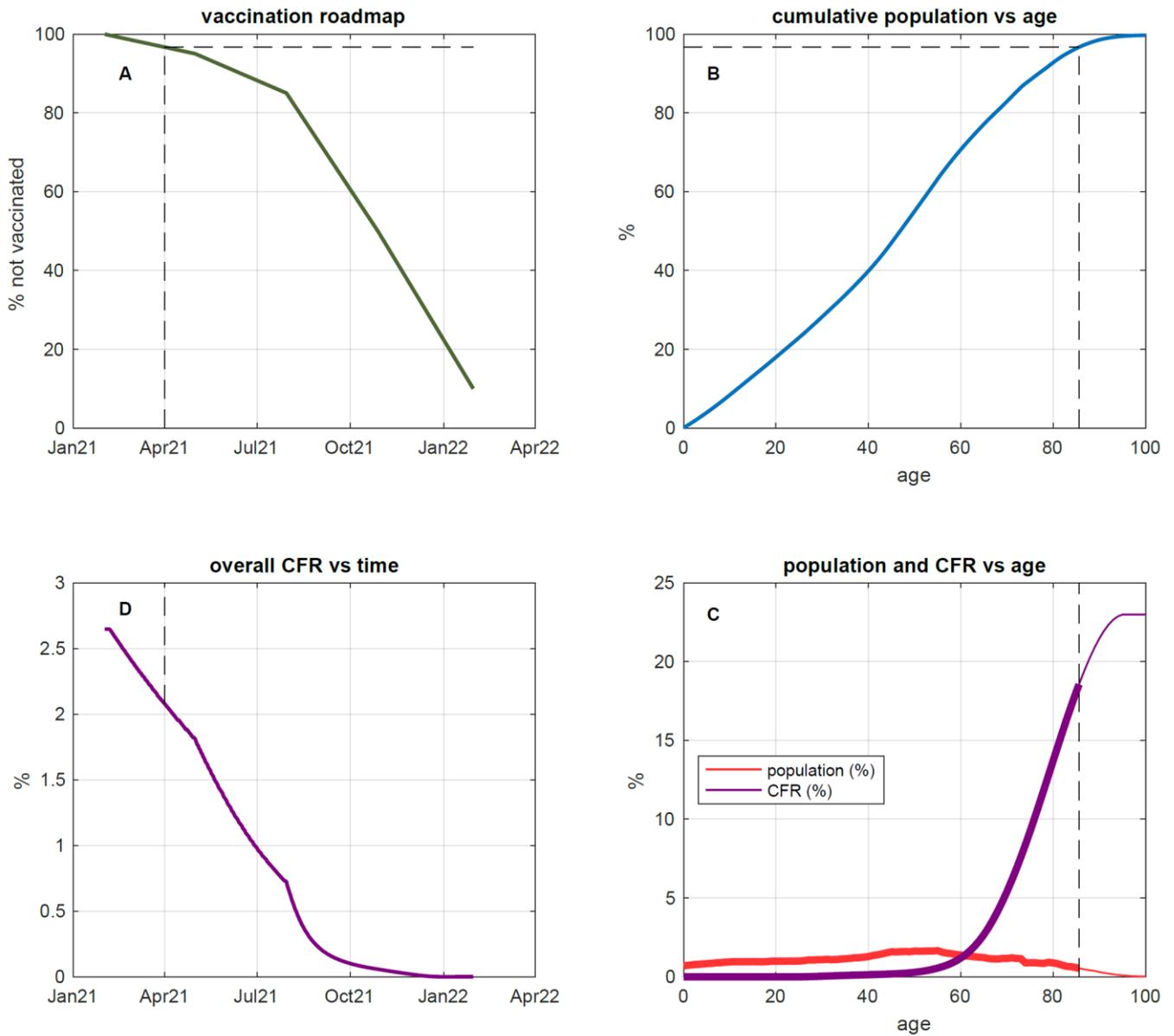

**Extended Data Figure 5.** Computation of the time-varying Case Fatality Rate as a function of the vaccination roadmap, taking population age classes into account. **A:** Considered vaccination roadmap: fraction of non-vaccinated (non-immunised) population over time. **B:** Cumulative Italian population as a function of age. **C:** Fraction of the population (red) and Case Fatality Rate (purple) as a function of age. **D:** Evolution of the overall Case Fatality Rate over time, assuming a vaccination schedule that gives priority to the elderly.

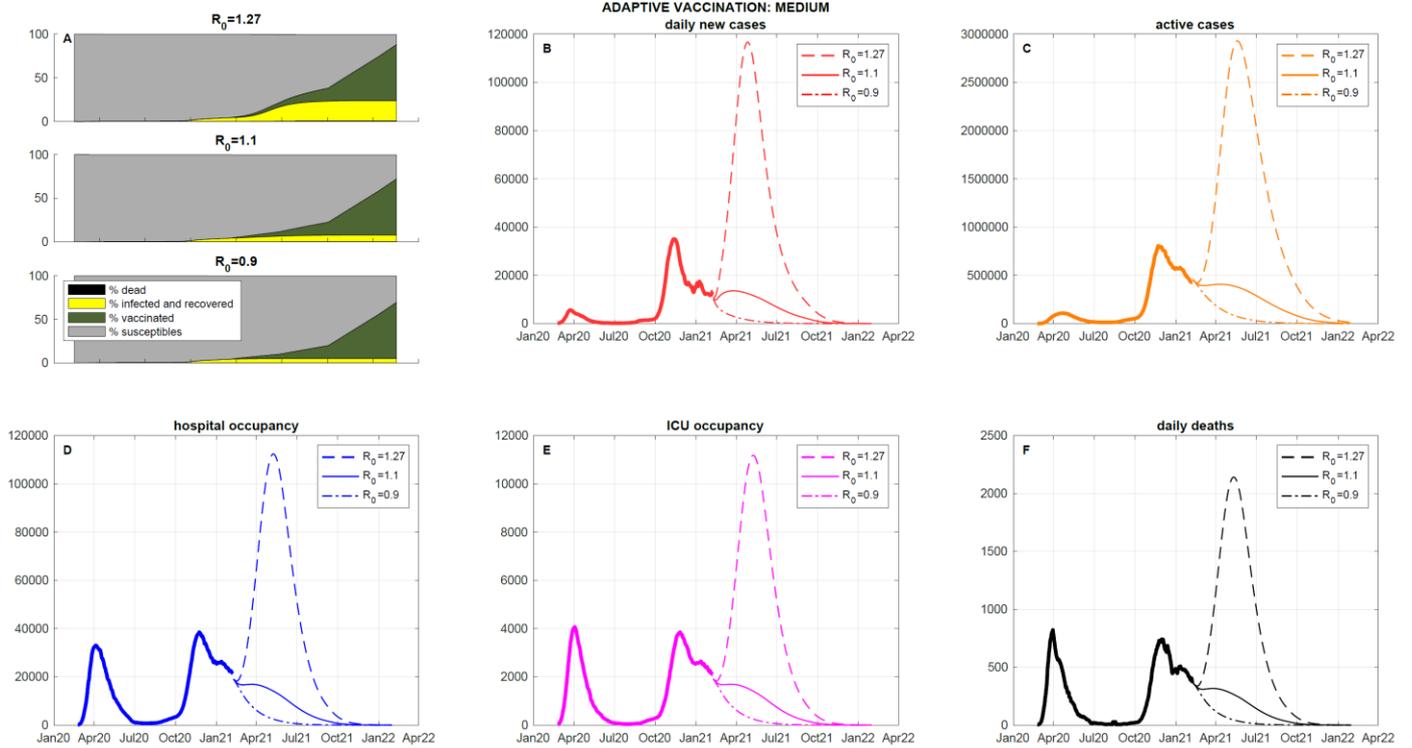

**Extended Data Figure 6.** Time evolution of the epidemic, with adaptive vaccination, when different constant values of $\mathcal{R}_0$, namely $\mathcal{R}_0 = 1.27$ (dashed line), $\mathcal{R}_0 = 1.1$ (solid line), $\mathcal{R}_0 = 0.9$ (dash-dotted line), are assumed, associated with different variants and/or containment strategies. The adaptive vaccination speed is inversely proportional to the size of the epidemic, i.e. to the number of infection cases. **A:** Time evolution of the fractions of susceptibles, infected and recovered, and dead. **B:** Daily new cases. **C:** Active cases. **D:** Hospital occupancy. **E:** ICU occupancy. **F:** Daily deaths.

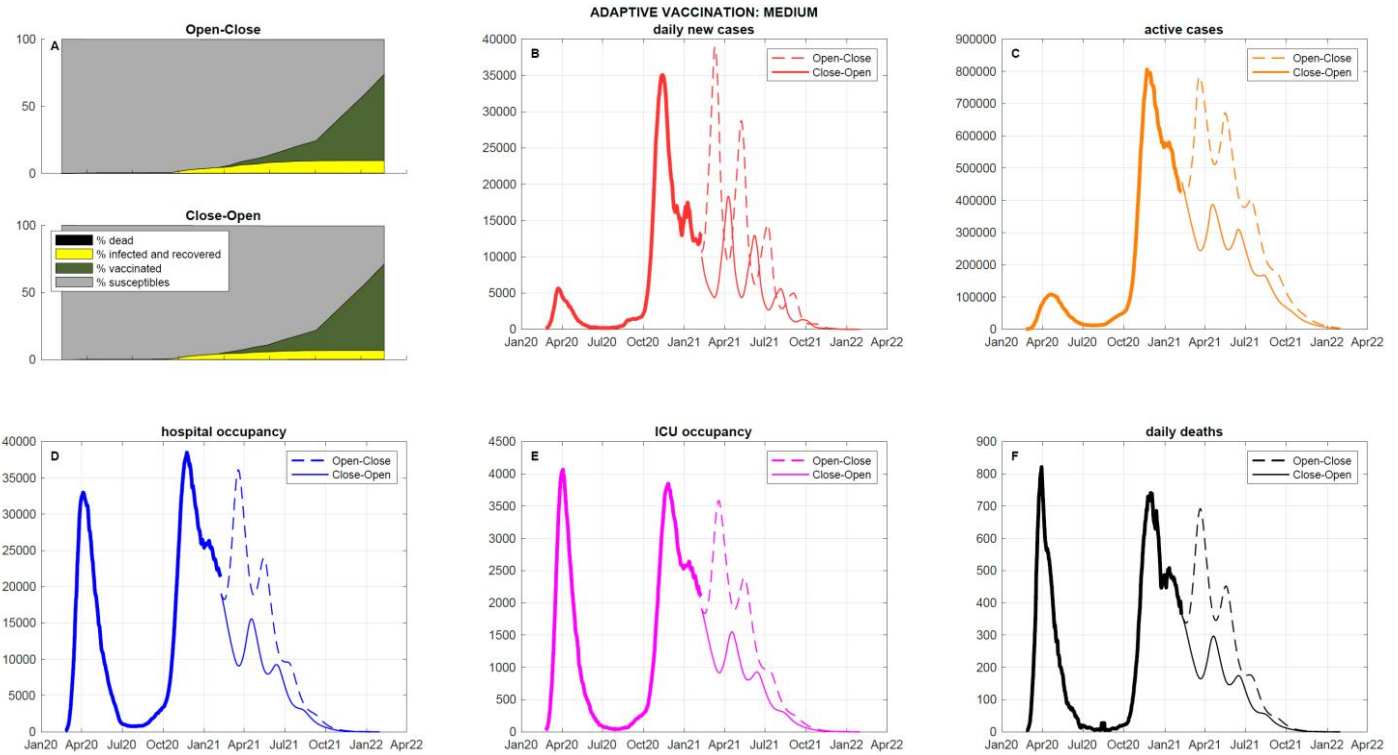

**Extended Data Figure 7.** Time evolution of the epidemic, in the absence of vaccination, when different intermittent strategies are enforced, with an average value of $\mathcal{R}_0$ equal to 1.1. The Open-Close strategy (dashed line) switches every month between $\mathcal{R}_0 = 1.27$ and $\mathcal{R}_0 = 0.9$, starting with $\mathcal{R}_0 = 1.27$. The Close-Open strategy (solid line) does the same, but starts with $\mathcal{R}_0 = 0.9$. The adaptive vaccination speed is inversely proportional to the size of the epidemic, i.e. to the number of infection cases. **A:** Time evolution of the fractions of susceptibles, infected and recovered, and dead. **B:** Daily new cases. **C:** Active cases. **D:** Hospital occupancy. **E:** ICU occupancy. **F:** Daily deaths.

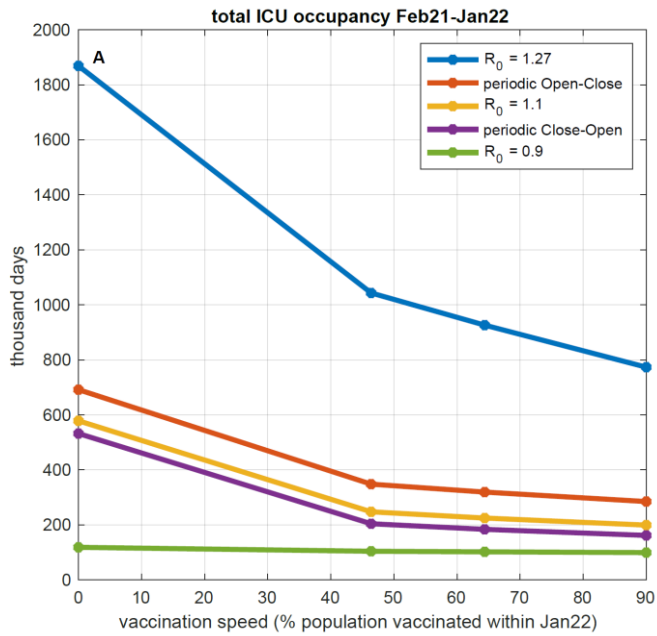
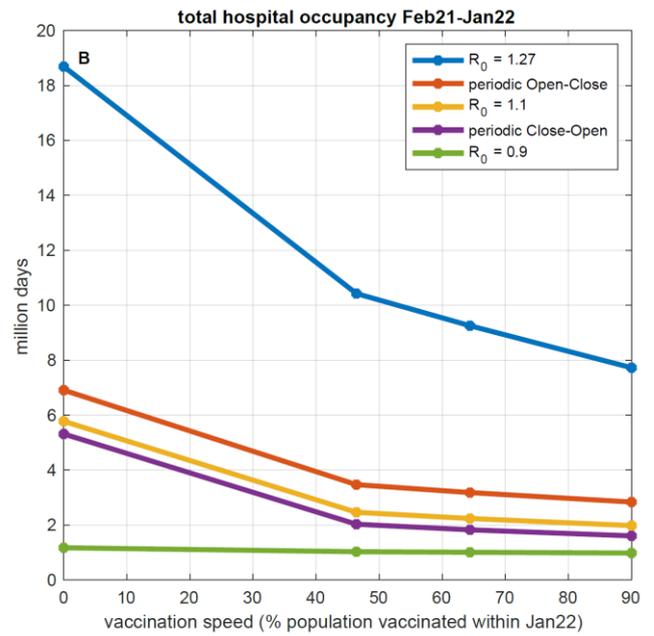

**Extended Data Figure 8.** Vaccination-cost curves: for a given $\mathcal{R}_0$ profile, the curve gives the total ICU occupancy (**A**) and the total hospital occupancy (**B**) in the period from February 2021 to January 2022, as a function of the average vaccination speed, measured as the fraction of vaccinated population at the end of the period. The vaccination-cost curves corresponding to a constant reproduction number are reported in blue ($\mathcal{R}_0 = 1.27$), yellow ($\mathcal{R}_0 = 1.1$) and green ($\mathcal{R}_0 = 0.9$), while those corresponding to intermittent strategies are reported in orange (Open-Close) and purple (Close-Open).

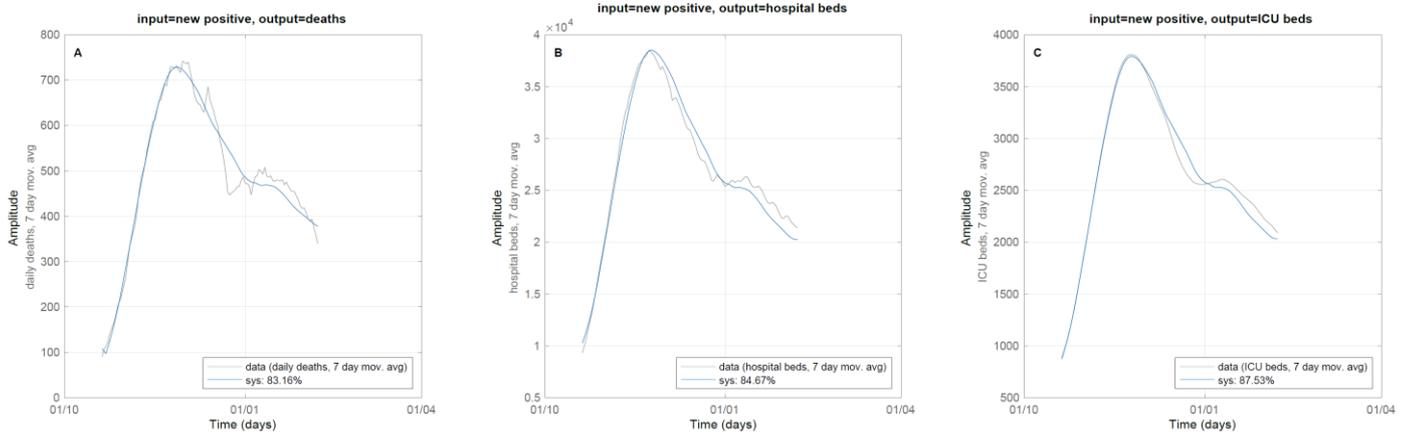

**Extended Figure 9.** Estimation of the three dynamic input-output models linking new cases (input) to three outputs: deaths (**A**), hospital occupancy (**B**) and ICU occupancy (**C**). Each panel displays the observed outputs (grey) and the values predicted by the identified dynamic system (blue). In order to allow for weekly oscillations both input and output series were prefiltered with a 7-day moving average before performing nonlinear least squares fitting of a first-order model with delay. The FIT ratios of the three models are 83.16%, 84.67%, and 87.53%.

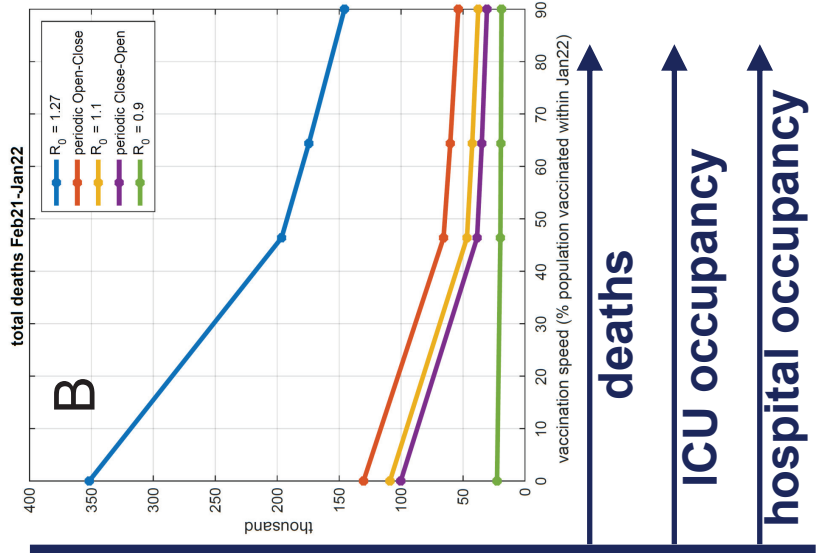
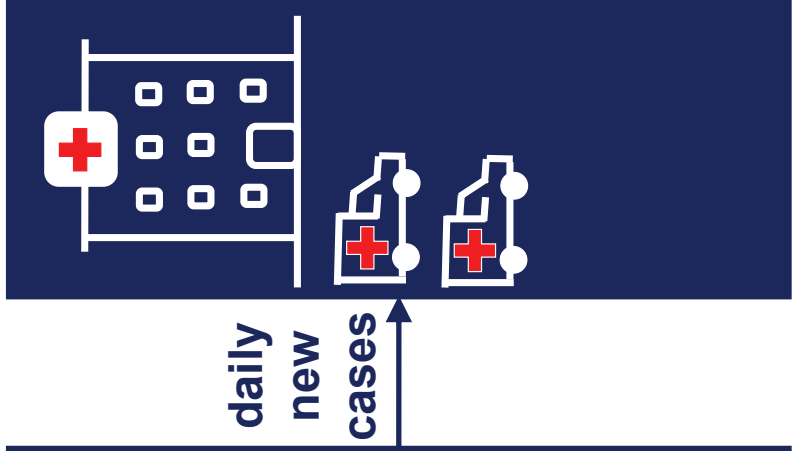
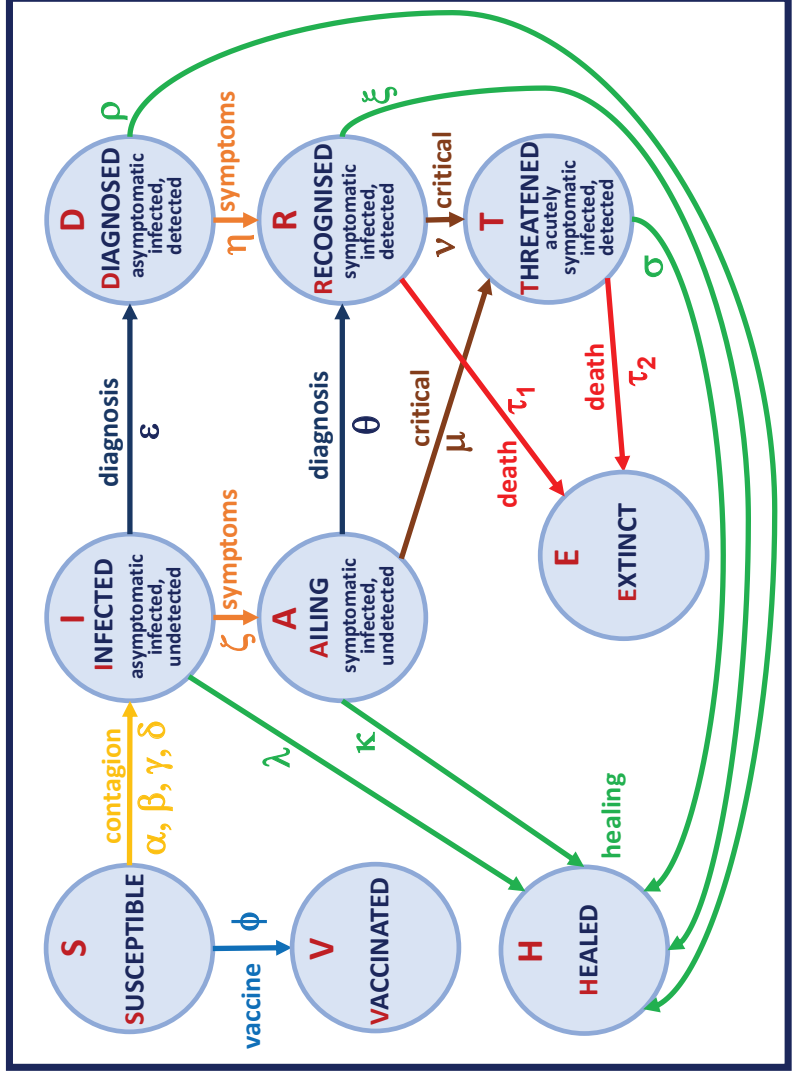

## VACCINATION: MEDIUM SPEED

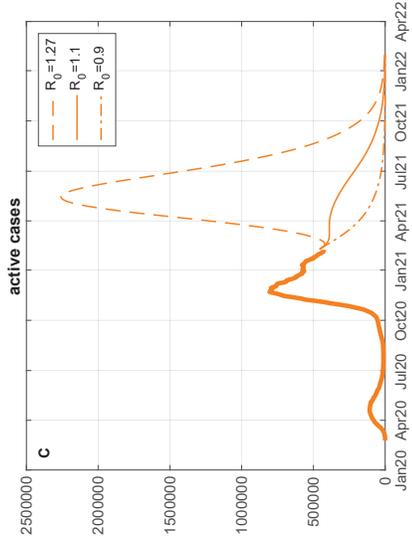
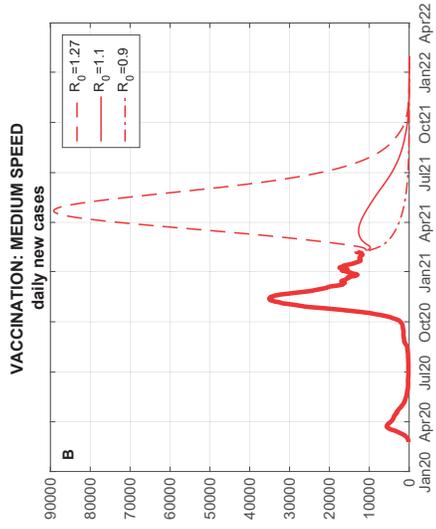
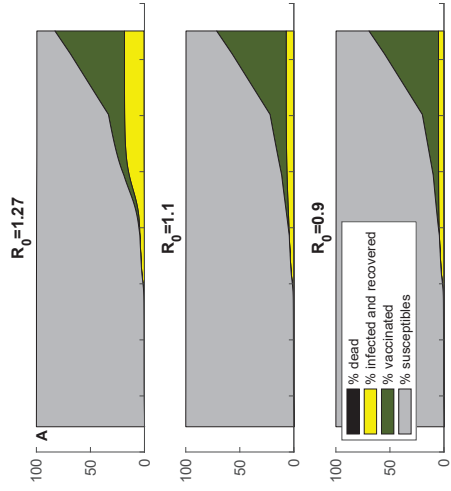
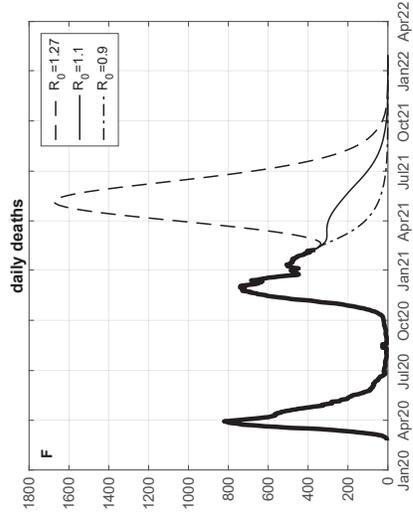
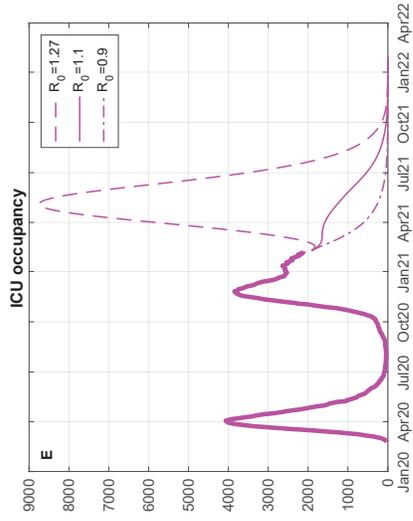
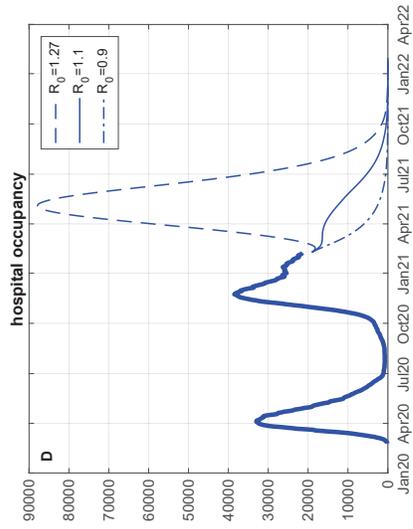

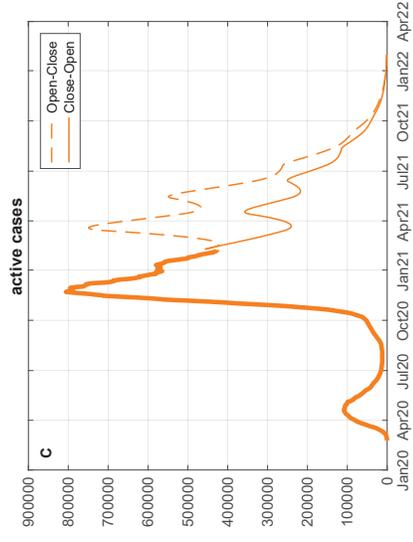
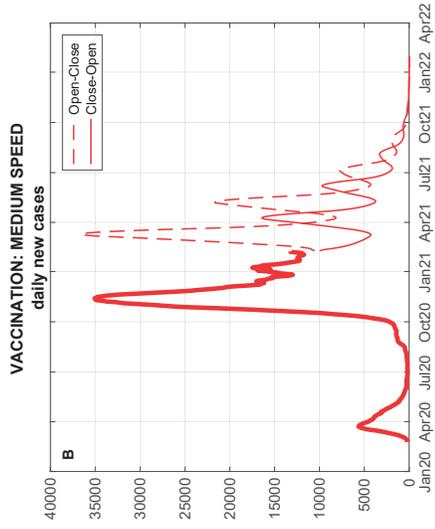
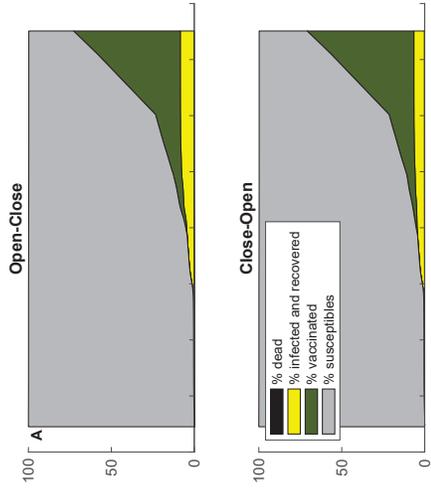
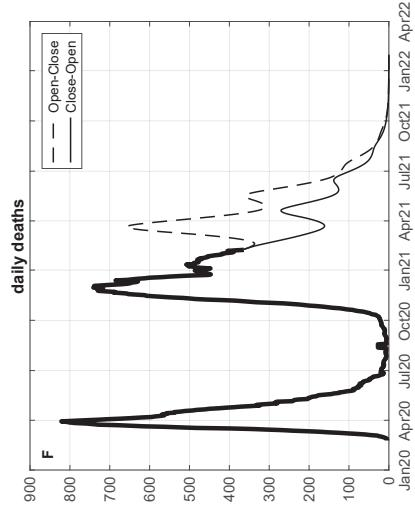
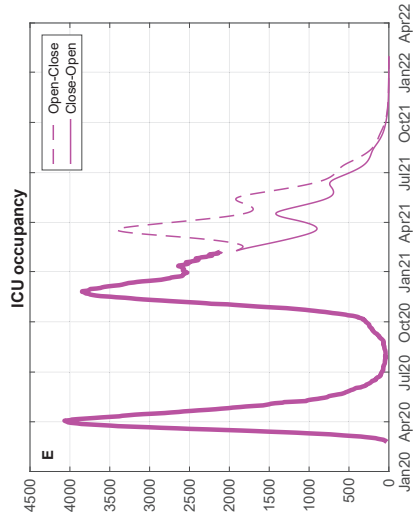
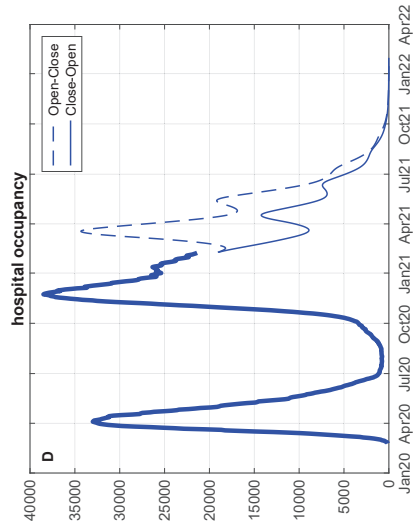

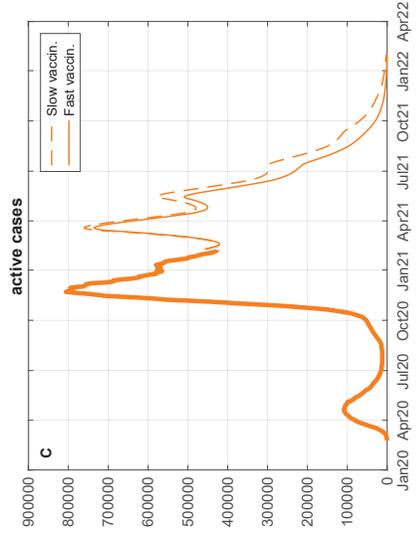
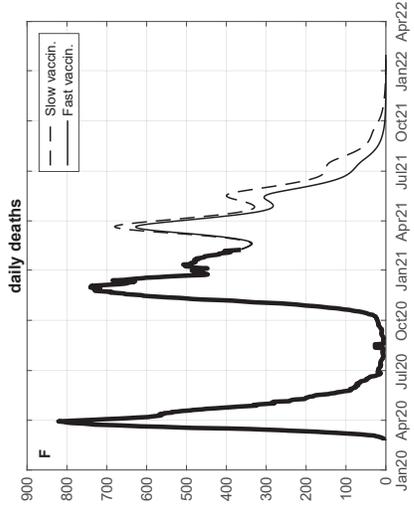
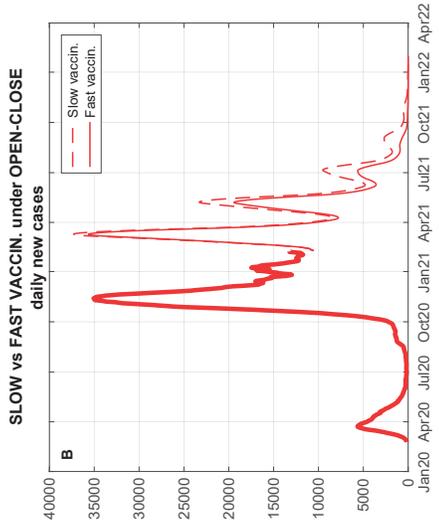
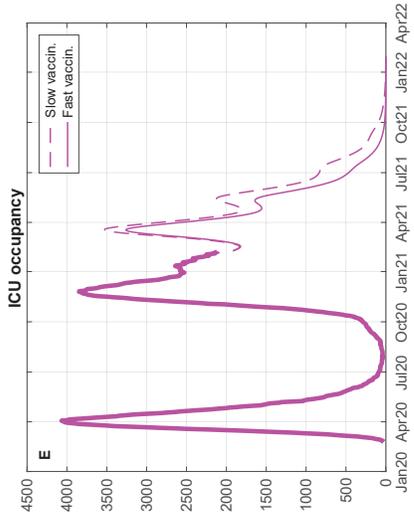
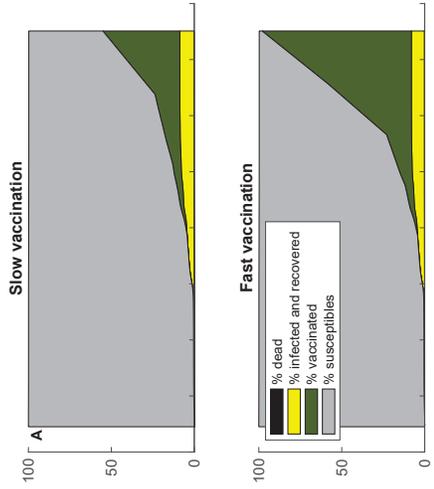
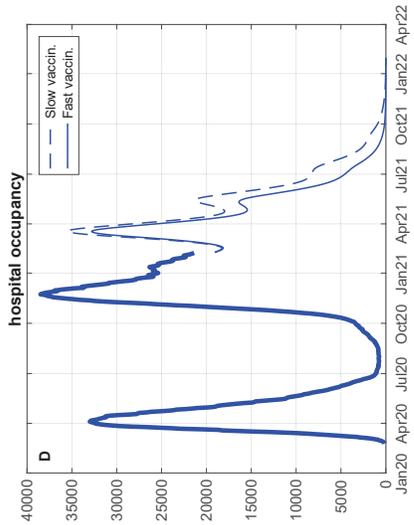

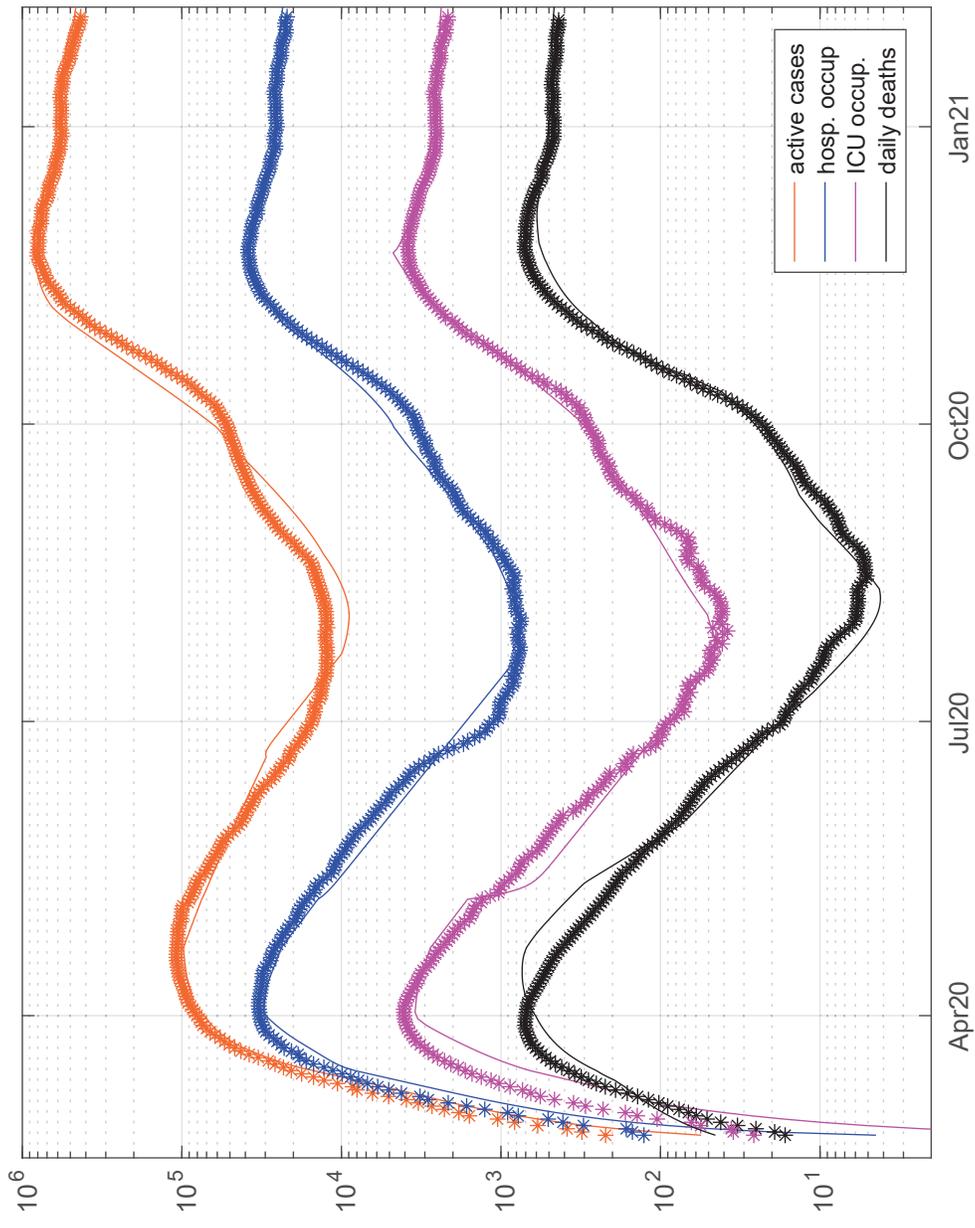

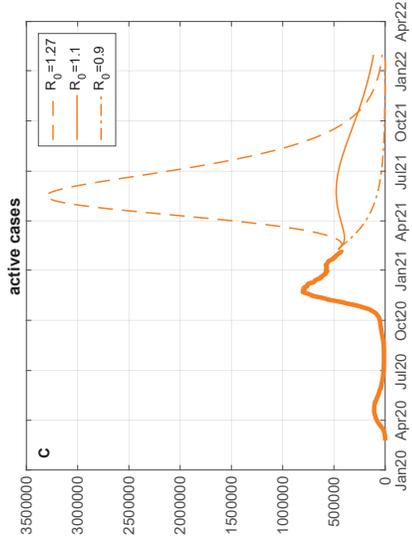
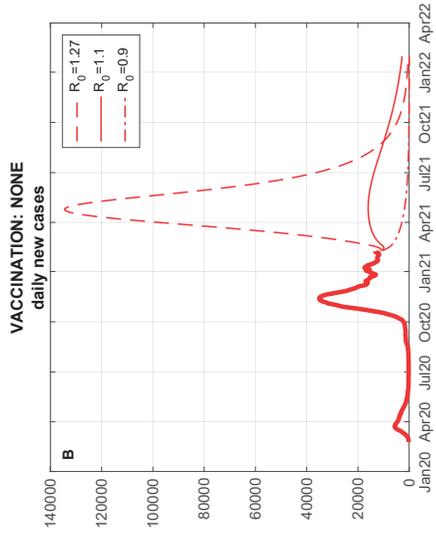
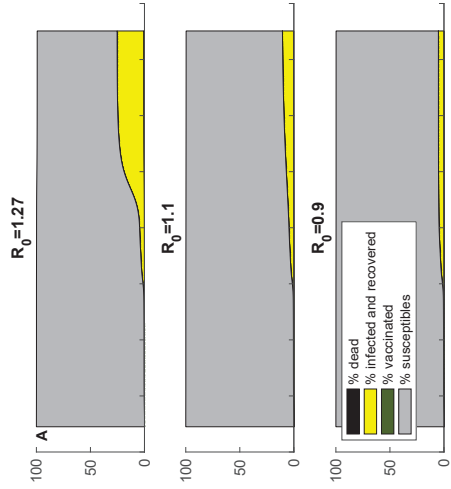
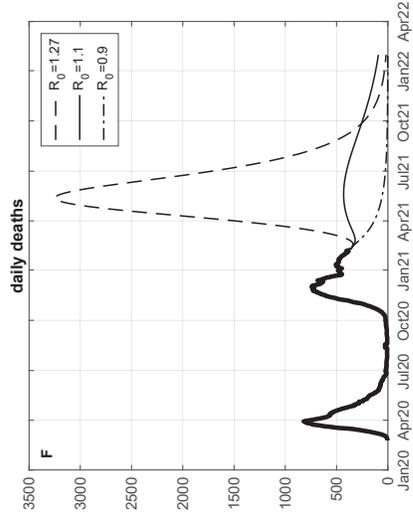
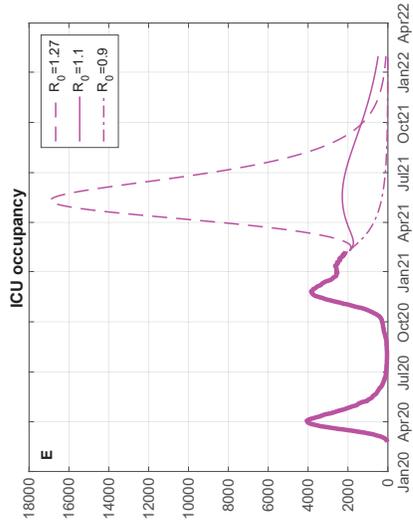
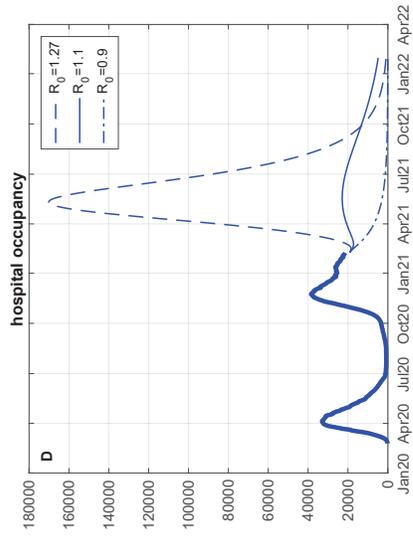

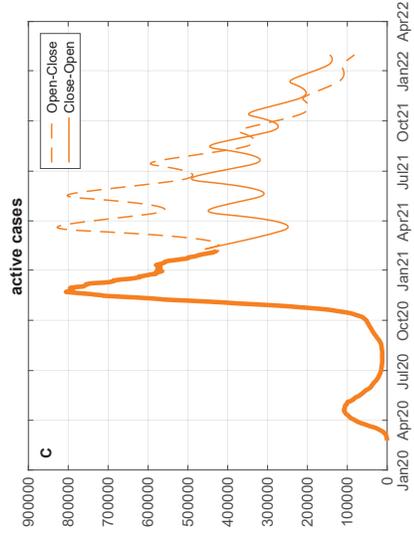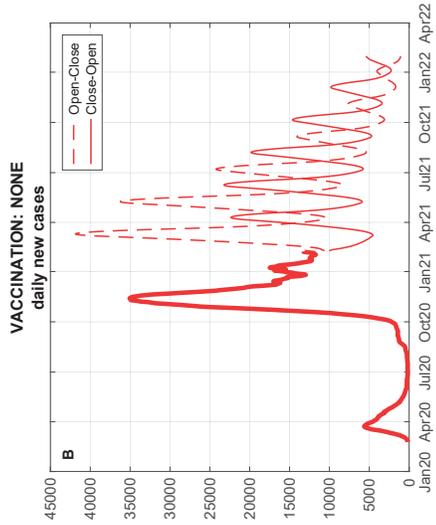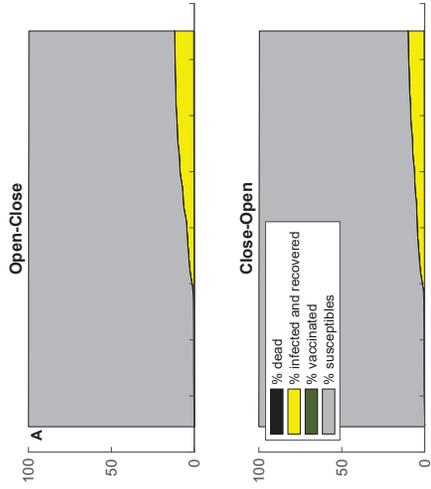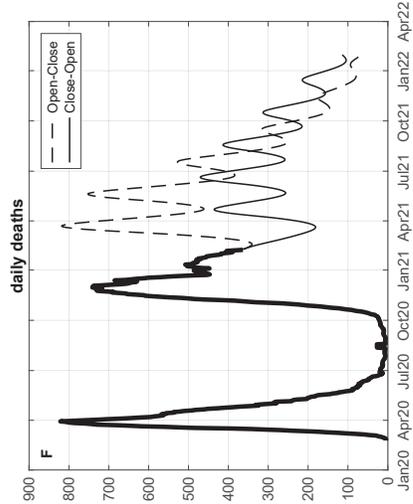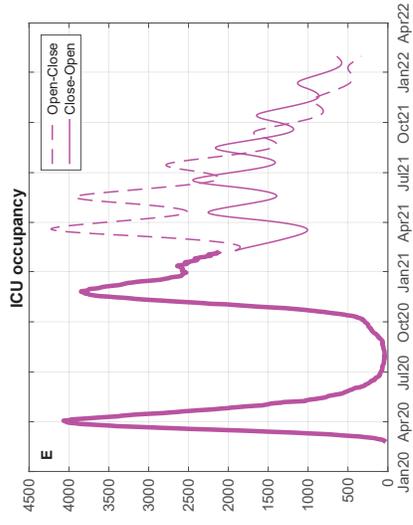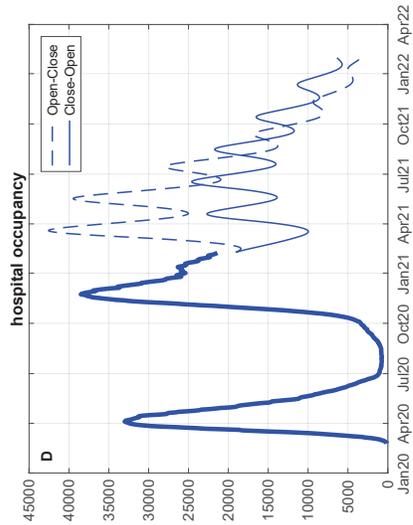

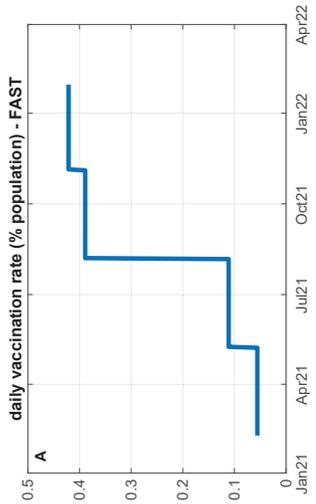
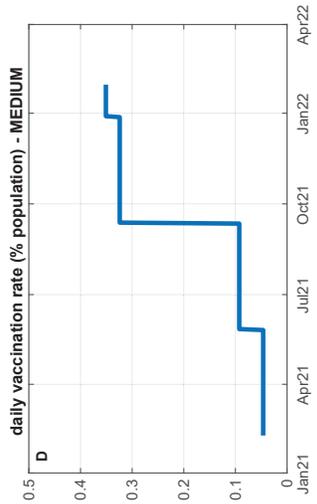
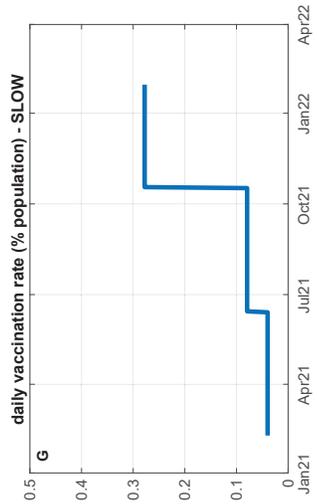
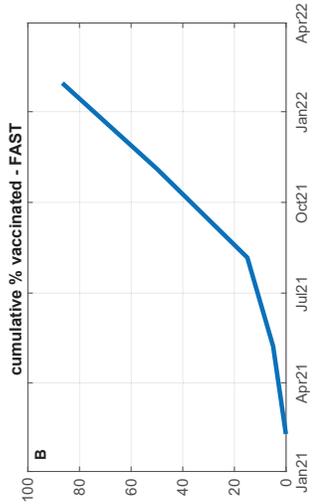
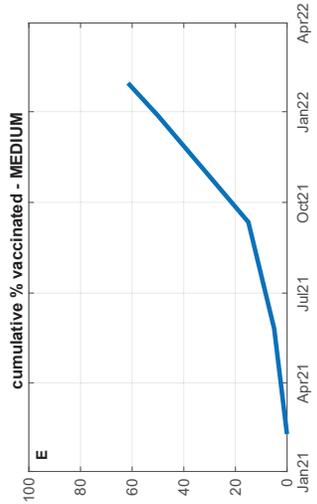
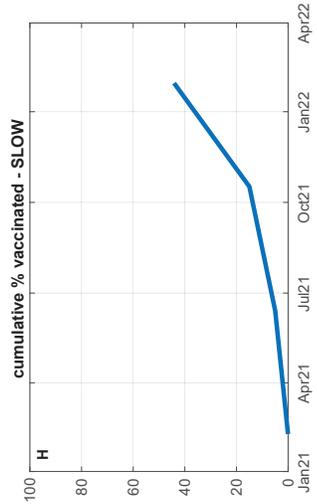
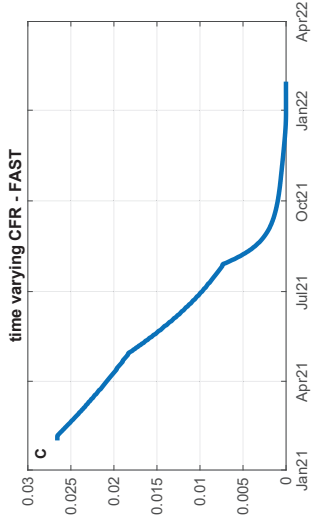
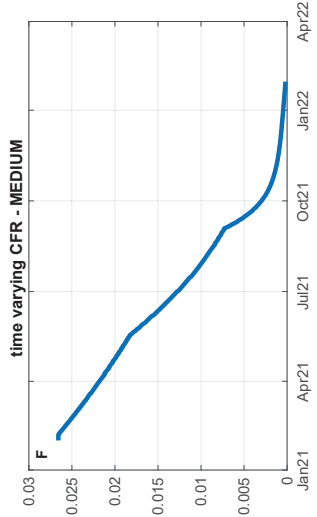
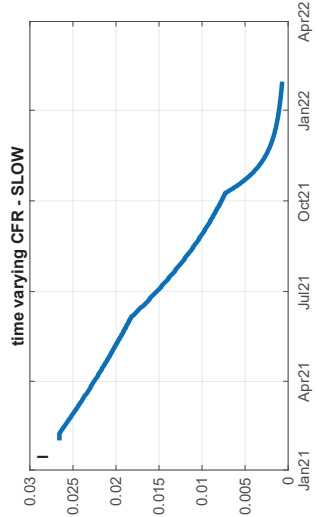

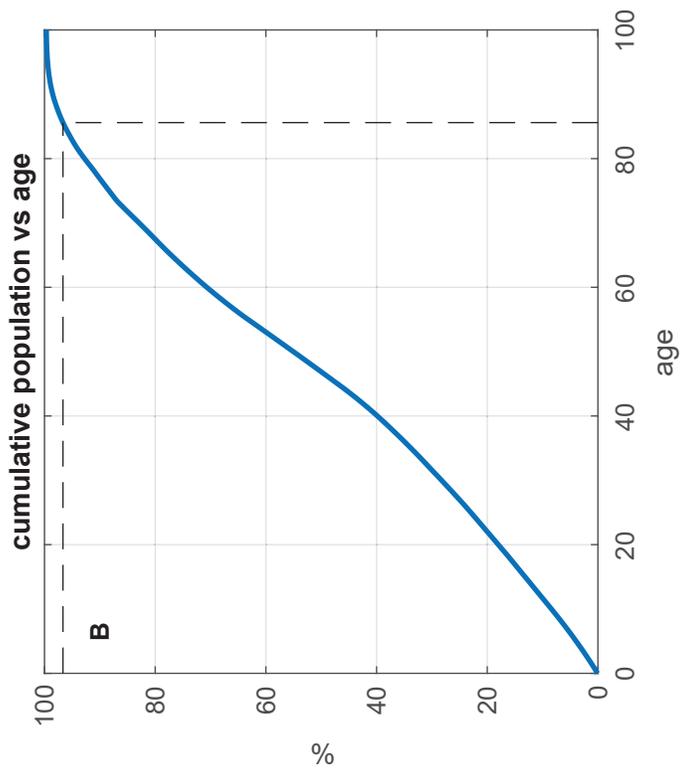
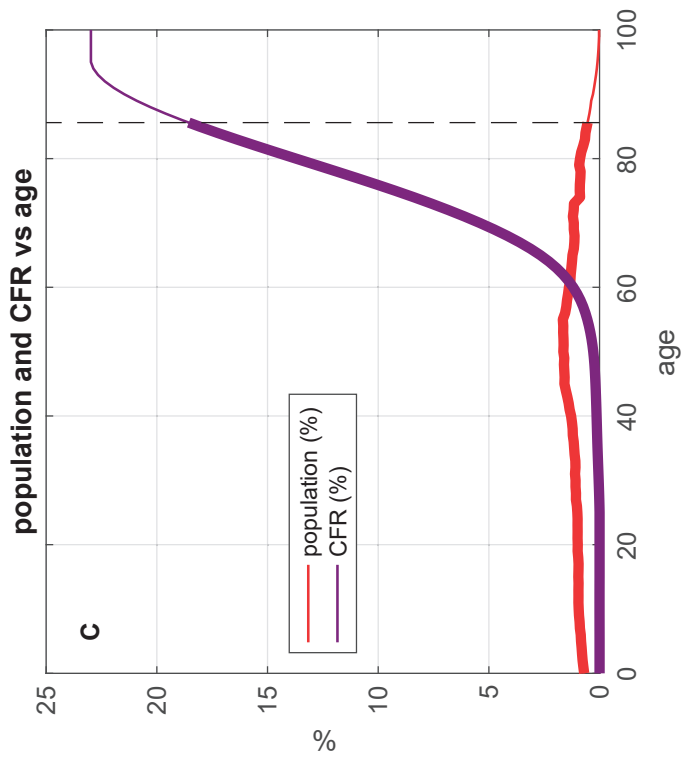
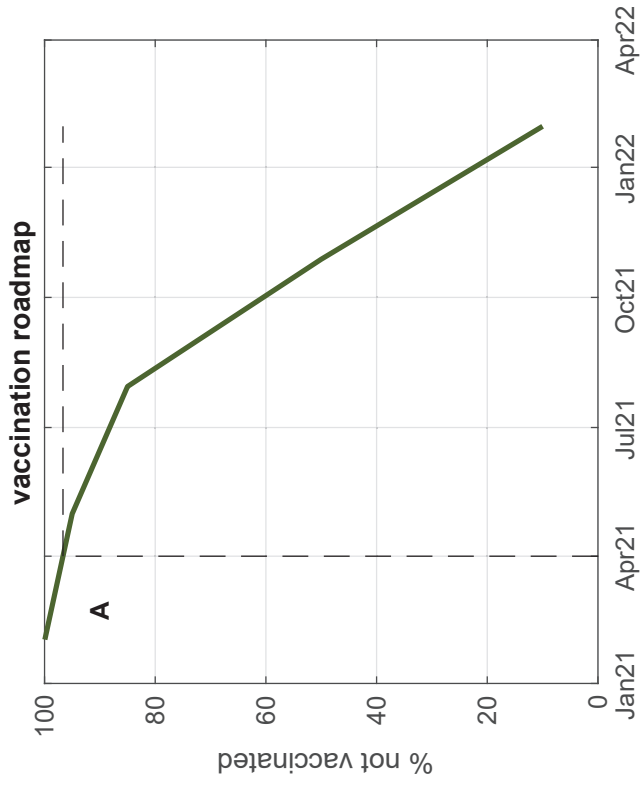
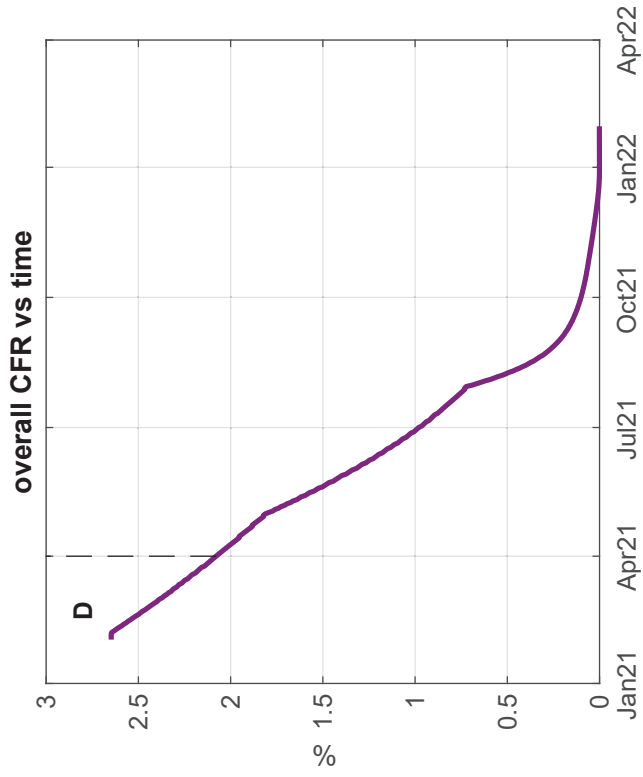

## ADAPTIVE VACCINATION: MEDIUM

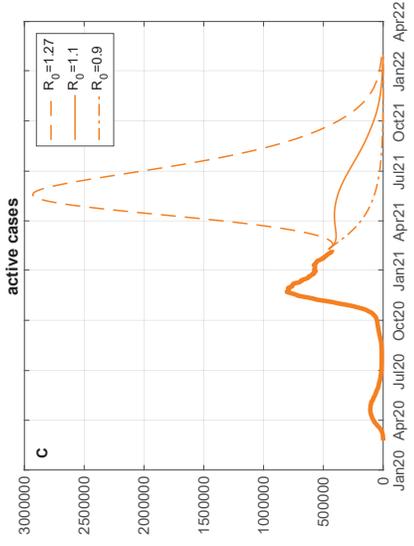

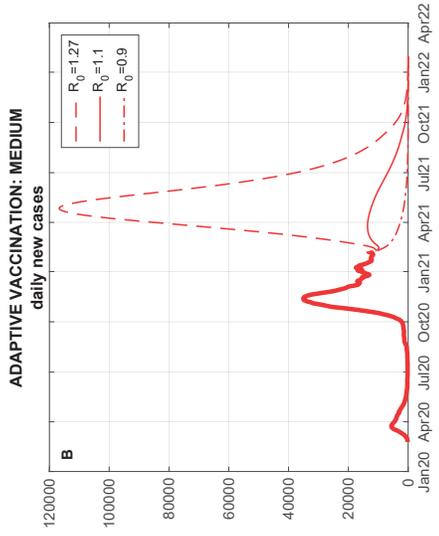

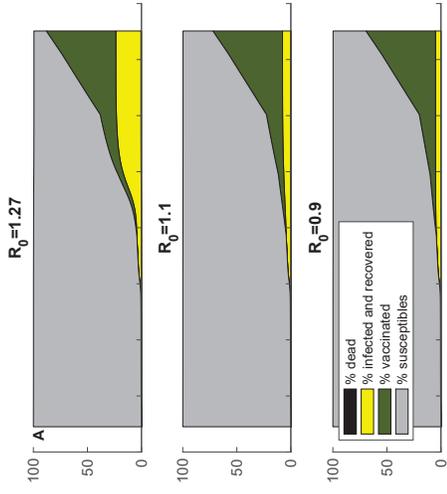

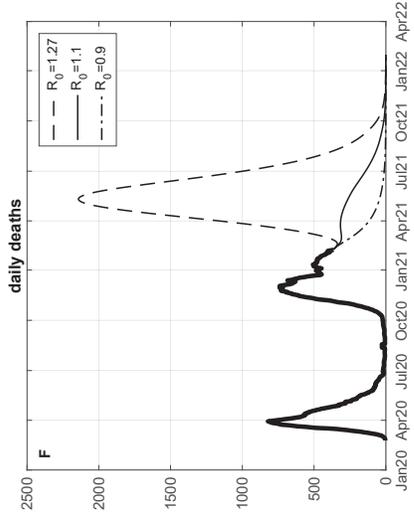

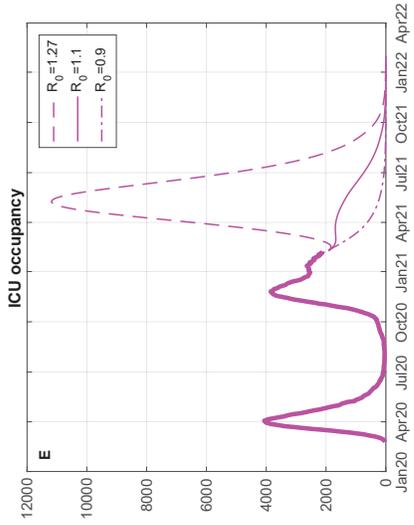

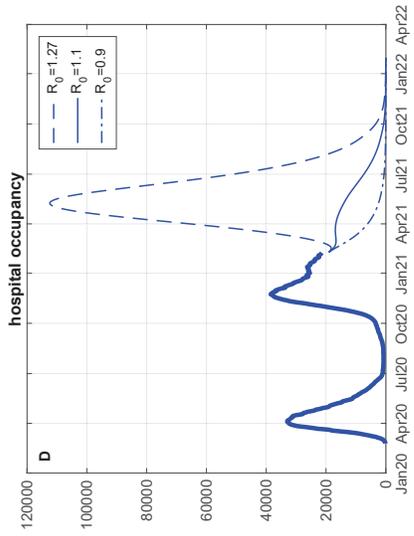

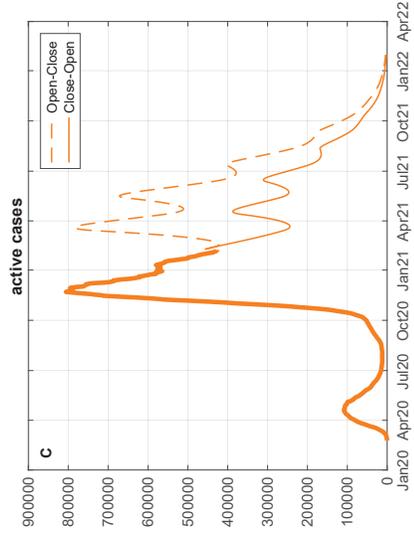
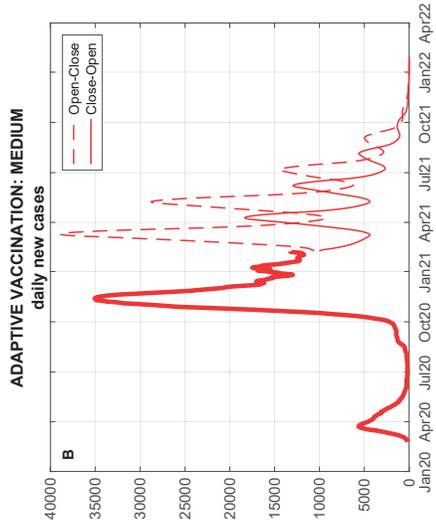
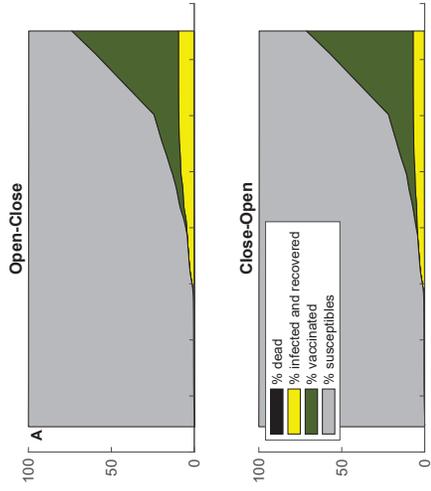
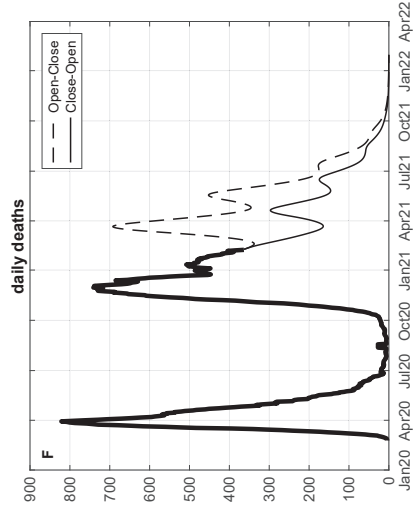
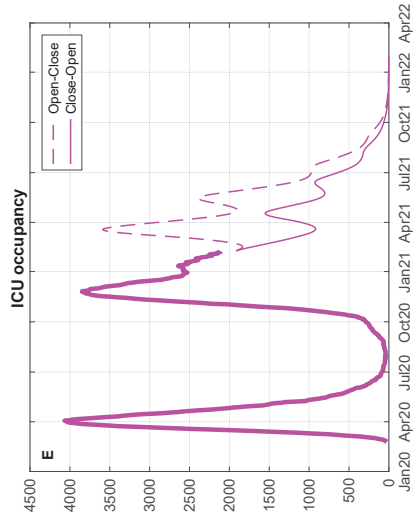
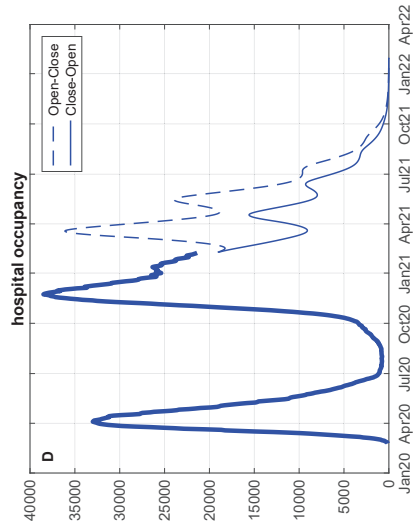

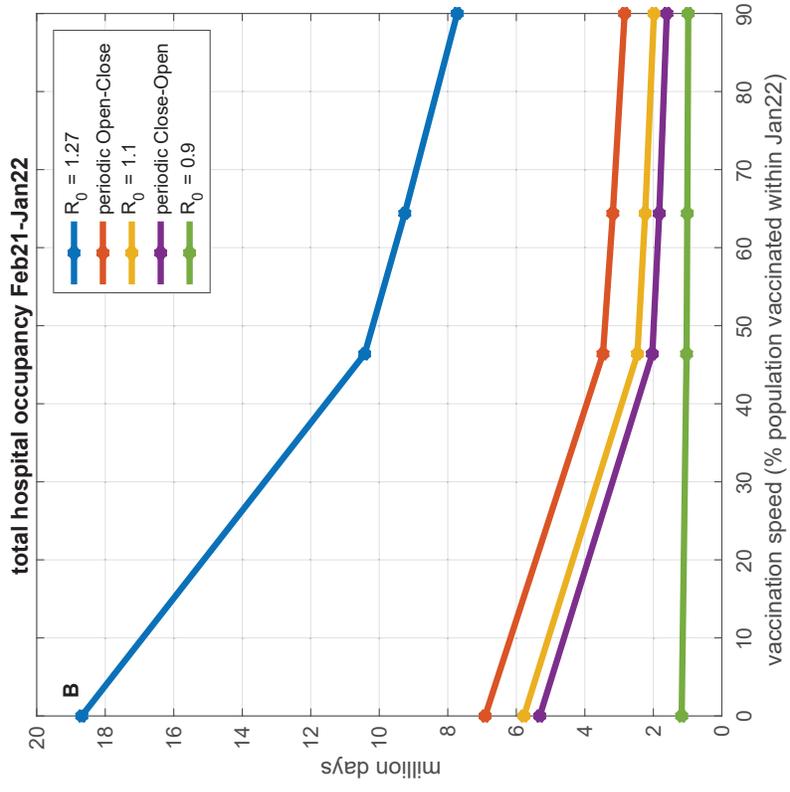

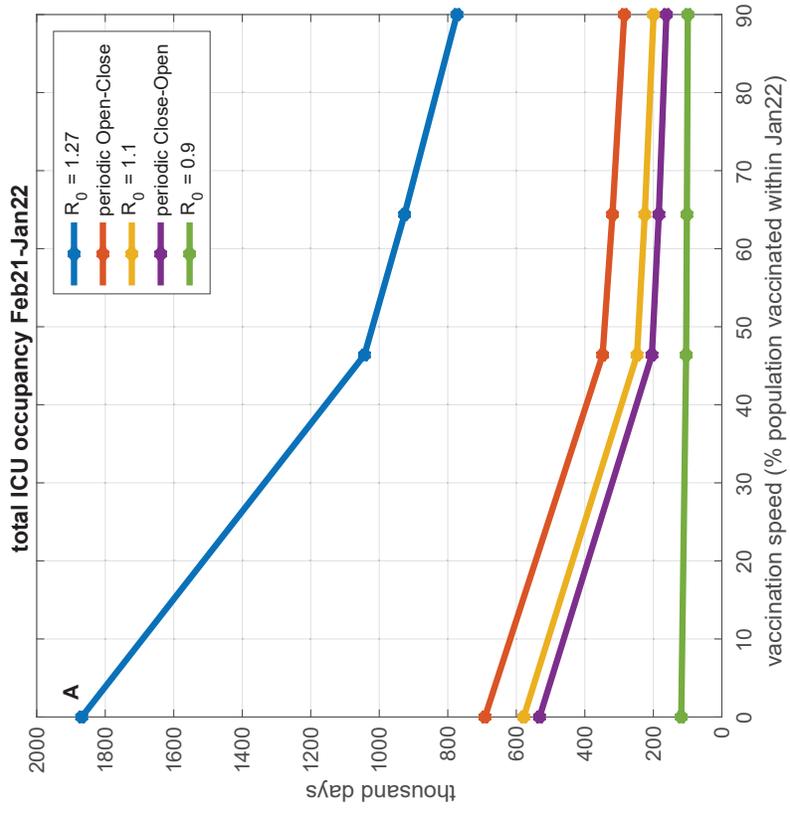

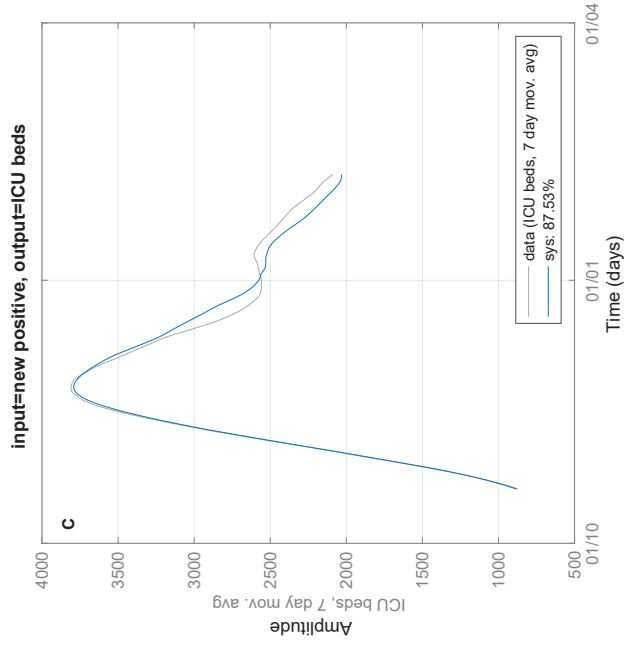

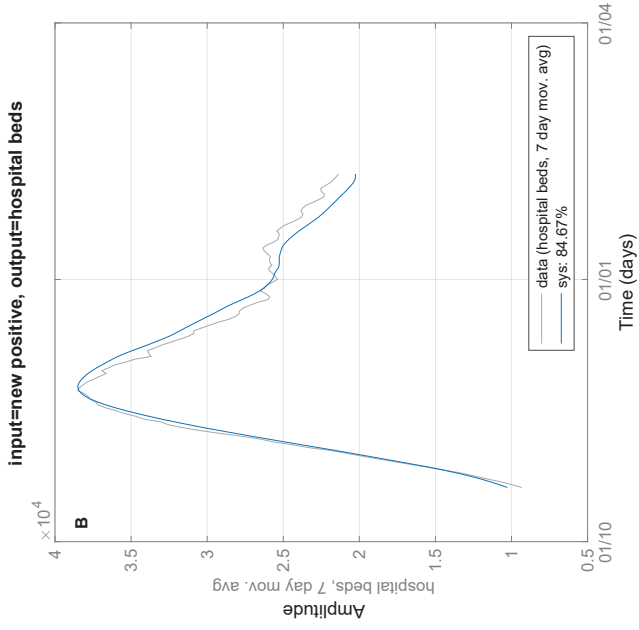

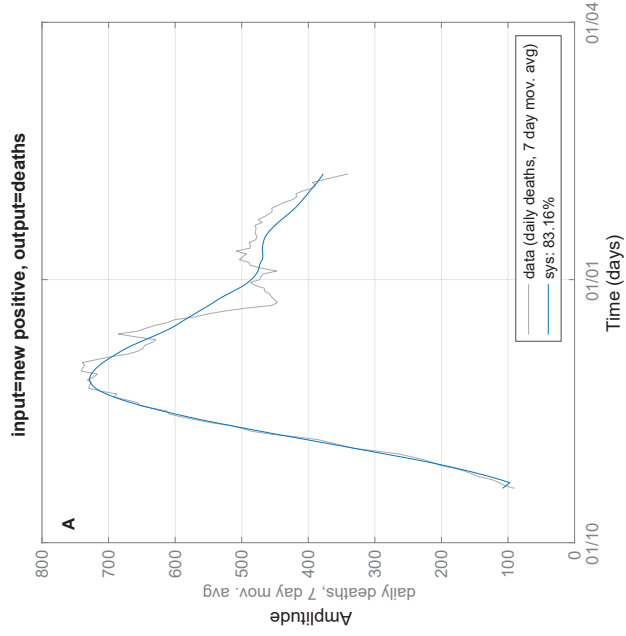